# Inductive-Effect-Driven Tunability of Magnetism and Luminescence in Triangular Layers ANd(SO$_4$)$_2$ (A = Rb, Cs)

Xudong Huai,[1] Ebube E. Oyeka,[1] Uchenna Chinaegbomkpa,[1] Michał J. Winiarski,[2] Hugo Sanabria,[3] and Thao T. Tran*[1]

AUTHOR ADDRESS

1. Department of Chemistry, Clemson University, Clemson, South Carolina 29634, United States
2. Faculty of Applied Physics and Mathematics and Advanced Materials Center, Gdansk University of Technology, Narutowicza 11/12, 80-233 Gdansk, Poland
3. Department of Physics and Astronomy, Clemson University, Clemson, South Carolina 29634, United States

**ABSTRACT:** Tuning the energy landscape of many-body electronic states in extended solids through the inductive effect—a concept widely used in organic chemistry—offers a new, effective strategy for materials development. Here, we demonstrate this approach using the ANd(SO$_4$)$_2$ (A = Rb, Cs) model system, which possesses different A-site electronegativity and displays a distorted triangular lattice of Nd$^{3+}$ ($^4I_{9/2}$ ground term). Magnetization data indicate appreciable antiferromagnetic interactions without long-range ordering down to 1.8 K while highlighting the tunable population of the electronic states. Temperature-dependent and time-resolved photoluminescence measurements reveal that emissions and nonradiative processes can be modified by the inductive effect at the atomic level. Heat capacity data confirm no magnetic ordering and add insight into the role of phonons in emission lifetime. Density functional theory calculations prove enhanced covalency in the Cs compound compared to the Rb counterpart while acknowledging the adjustable magnetic intralayer and interlayer exchange pathways. These results demonstrate a viable framework for utilizing the inductive effect as an important knob for simultaneously dialing in magnetic, optical, and electronic properties in quantum materials.

## INTRODUCTION

In frustrated magnets, various electronic states either share similar energy levels or are separated by small energy barriers.[1-7] This complex energy landscape results in distinctive spin configurations and unusual dynamics.[8-16] However, tuning this energy space and thus the physical behaviors of quantum materials remains a significant challenge. The inductive effect, which is largely developed in organic chemistry to modulate covalency in molecules, offers a promising pathway to tune new spin physics, phonon-assisted emission, and quantum functions in extended solids.[17-23] To implement this strategy, we designed and synthesized two new frustrated magnets, ANd(SO$_4$)$_2$ (A = Rb, Cs), which feature a distorted Nd triangular magnetic lattice. Despite being in the same group (group 1A), Rb and Cs differ in electronegativity, providing an opportunity to modulate material properties through the inductive effect while maintaining identical crystal structures and valence electrons. Nd$^{3+}$ was chosen for its non-integer spin $S$ = 3/2 with $^4I_{9/2}$ ground term and optically accessible spin states.

The interactions between Nd$^{3+}$ spins and variations in the electronegativity at the A site lead to modification in chemical bonds, and thus, the ligand field splitting effect (Figure 1). In this study, we comprehensively investigate the optical, magnetic, and thermomagnetic properties of ANd(SO$_4$)$_2$ through detailed temperature- and field-dependent measurements. The experimental results are complemented by density functional theory (DFT) calculations. Our goal is to elucidate the chemical origin of the physical properties of the triangular magnets ANd(SO$_4$)$_2$ and examine how the A-site choice influences their magnetism, phonons, and luminescence.

## RESULTS AND DISCUSSION

ANd(SO$_4$)$_2$ (A = Rb, Cs) crystalizes in an orthorhombic *P*nna space group and features a distorted 2D-triangular

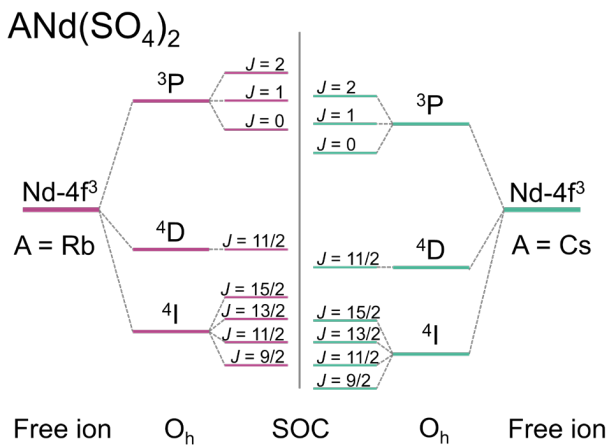

**Figure 1.** Proposed energy diagram of the electronic states in ANd(SO$_4$)$_2$ (A = Rb, Cs) in the presence of ligand field and spin-orbit coupling (SOC) showing the impact of the inductive effect introduced by the A-site.

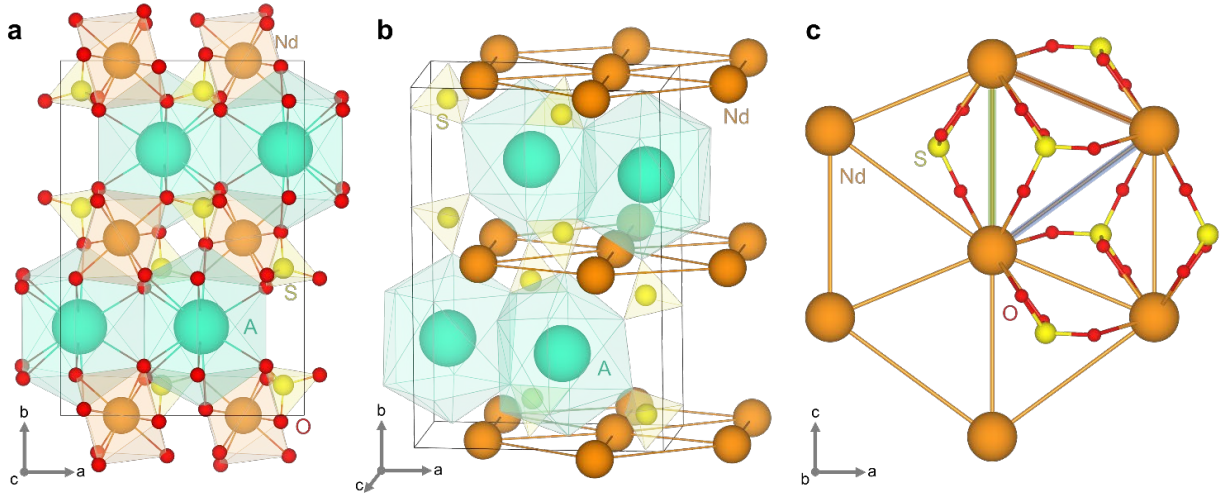

**Figure 2**. Crystal structure of (a) ANd(SO$_4$)$_2$ (A = Rb, Cs) and (b) the distorted Nd triangular layers in ANd(SO$_4$)$_2$. (c) A closer look at the triangular magnetic lattice showing 3 different interaction pathways between Nd$^{3+}$ ions.

Nd$^{3+}$ lattice (Figure 2). Each Nd$^{3+}$ ($S$ = 3/2, 4$f^3$) is coordinated by six oxygen atoms in a pseudo-octahedral geometry, with Nd-O bond lengths ranging from 2.438(1) Å to 2.555(1) Å. Within the Nd triangular layer, the Nd atoms are connected through SO$_4$ groups (Figure 2b), with Nd—Nd distances ranging from 5.099(1) to 5.911(1) Å. The distorted triangular geometry of Nd arises from different fashions by which the Nd ions are linked (Figure 2c): corner-sharing and/or edge-sharing through the SO$_4$ units.[10, 11, 24]

Temperature-dependent magnetization measurements were conducted at $\mu_0H$ = 0.1 T, and field-dependent magnetization was measured at $T$ = 1.8 K (Figure 3). Fitting the high-temperature paramagnetic data with a Curie-Weiss model yielded an effective magnetic moment $\mu_{eff}$ = 3.6(1) $\mu_B$ per Nd$^{3+}$ ion in RbNd(SO$_4$)$_2$, consistent with the expected ground-state spin-orbit coupling moment (3.62 $\mu B$, $J$ = 9/2, $g$ = 8/11).[25-27] In contrast, the fitted effective magnetic moment for CsNd(SO$_4$)$_2$ is $\mu_{eff}$ = 4.5(1) $\mu_B$ per Nd$^{3+}$ ion, which aligns with the spin-orbit coupling value for the first excited state (4.35 $\mu B$, $J$ = 11/2, $g$ = 8/11).[28-31] The larger $\mu_{eff}$ in the Cs compound can be attributed to more covalency in the Cs compound, which leads to more diffused Nd$^{3+}$ states, and the energy separation between the $^4I_{9/2}$ ground state and the $^4I_{11/2}$ excited state becomes comparable to $kT$. As a result, the first excited state ($^4I_{11/2}$) becomes thermally populated and orbitally mixed with the ground state at elevated temperatures (250 K and above), where the Curie-Weiss fitting is performed (Figure 3). In contrast, in the Rb compound, the states are more separated, resulting in a predominant population of the $^4I_{9/2}$ ground state. Performing a linear fit at lower temperatures (50 ≤ $T$ K ≤ 100) for CsNd(SO$_4$)$_2$ yields a fitted effective magnetic moment $\mu_{eff}$ = 4.0(1) $\mu_B$, approaching the expected value for a $J$ = 9/2 ground state. The negative Curie-Weiss temperatures ($\theta_{CW}$) of -31.7(1) K for RbNd(SO$_4$)$_2$ and -82.8(1) K for CsNd(SO$_4$)$_2$ indicate antiferromagnetic (AFM) interactions at high temperature. In Nd$^{3+}$ triangular lattice systems, the relationship between Nd–Nd spacing and magnetic ordering is summarized in Figure S5.[11, 24, 32-34] Materials having shorter Nd–Nd distances (3.5–4.4 Å) tend to undergo magnetic ordering due to increased competing interactions. In contrast, systems with longer Nd–Nd distances (5.1–6.2 Å) show greater $\theta_{CW}$ values and absence of magnetic ordering. ANd(SO$_4$)$_2$ falls in this category, with Nd–Nd distances between ~5.1 and 5.9 Å. The AFM interaction appears stronger in the Cs compound compared to Rb. This can be explained by the difference in the $\mu_{eff}$, that a larger magnetic dipole tends to result in a stronger magnetic interaction. But we also want to keep in mind that the increased overlap between Nd$^{3+}$ and SO$_4^{2-}$ units, which are the pathway of the exchange interactions, can also lead to a stronger magnetic interaction in the triangular layer and a larger |$\theta_{CW}$| in the Cs compound. The field-dependent magnetization data (Figure 3c) at $T$ = 1.8 K deviates from ideal paramagnetic behavior as predicted by the Brillouin function ($J$ = 9/2, $g$ = 8/11). The relatively small magnetic moment is comparable to reported cases with Nd$^{3+}$ on a triangular lattice.[35, 36] The significant departure from the Brillouin function confirms the presence of sizable AFM interactions.

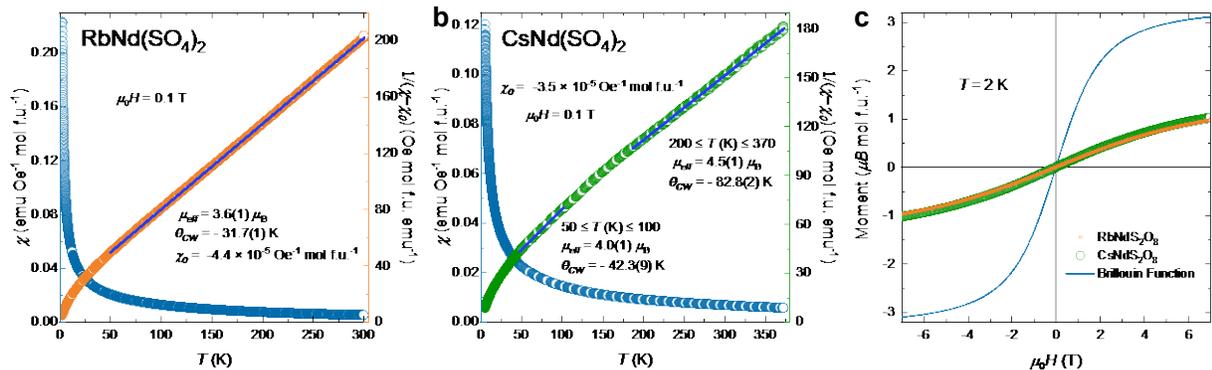

**Figure 3**. (a, b) Temperature dependent magnetic susceptibility with $\mu_0H$ = 0.1 T (blue) and Curie-Weiss analysis. (c). $M(H)$ curves at $T$ = 1.8 K and Brillouin function ($J$ = 9/2, $g$ = 8/11).

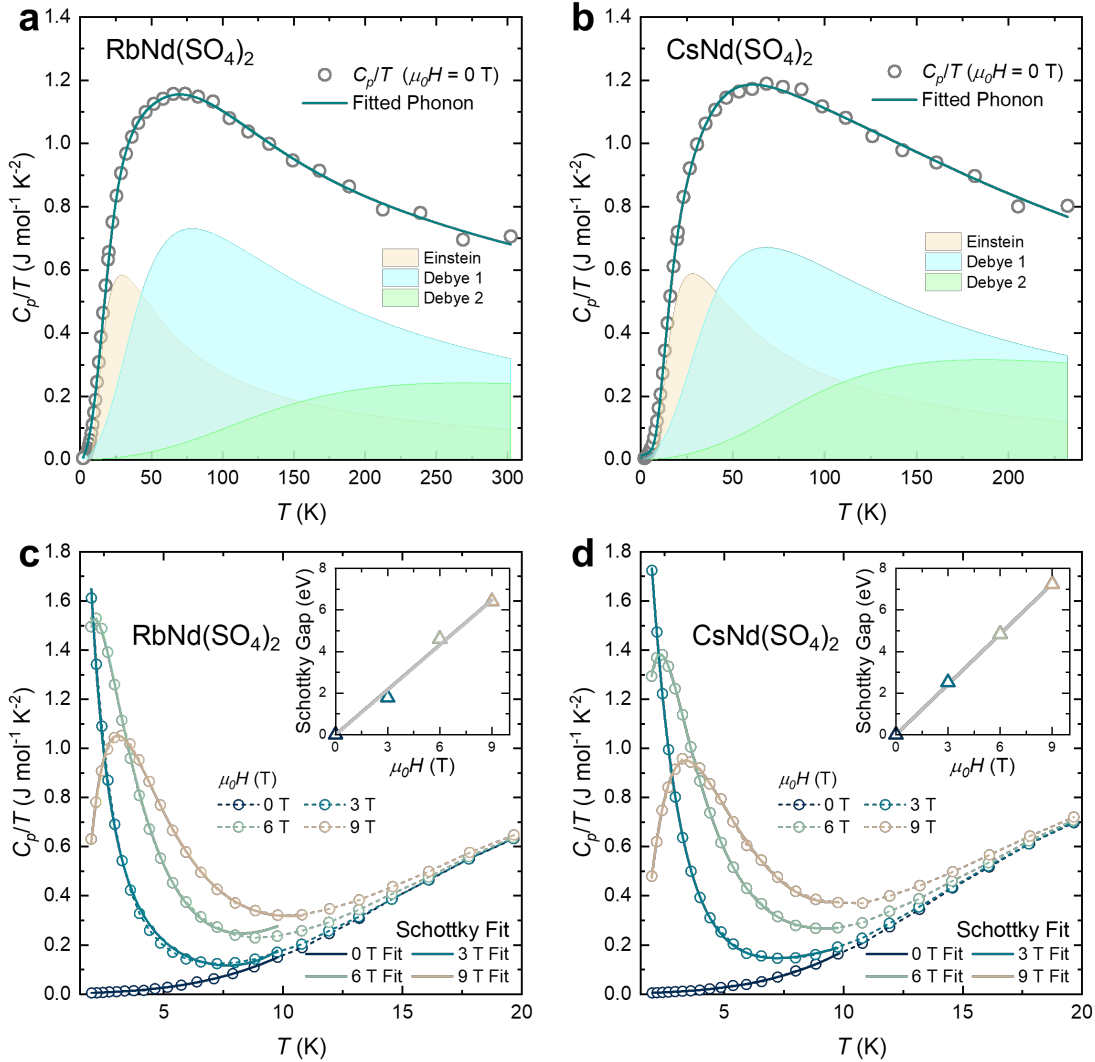

**Figure 4**. (a, b) Molar heat capacity over temperature ($C_p/T$) vs. temperature for ANd(SO$_4$)$_2$ at $\mu_0H = 0$ T and calculated phonon. (c, d) Schottky heat capacity under different fields with the insert showing the fitted Schottky gap with a two-level Schottky model.

To examine the thermomagnetic properties of ANd(SO$_4$)$_2$, the zero-field heat capacity measurements were conducted between 2 K and 300 K (Figure 4a, b). Consistent with the magnetization data, no anomaly is observed in the specific heat, indicating the absence of long-range magnetic ordering. To gain deeper insights into the phonon behavior, a phonon model was constructed to describe the specific heat data. This model includes one Einstein mode and two Debye modes (Equations 1 to 3) and is described by the following equations:

$$C_{Einstein} = 3NRs_E \frac{(\theta_E/T)^2 \exp(\theta_E/T)}{[\exp(\theta_E/T)-1]^2} \quad (1)$$

$$C_{Debye} = 9NRs_D \left(\frac{T}{\theta_D}\right)^3 D(\theta_D/T) \quad (2)$$

$$\frac{C_p}{T} = \frac{C_{Einstein(1)}}{T} + \frac{C_{Debye(1)}}{T} + \frac{C_{Debye(2)}}{T} \quad (3)$$

where $N$ is the number of atoms, $R$ is the gas constant, $s_E$ is the number of oscillators in the Einstein mode, and $\theta_E$ is the Einstein temperature, $k$ is Boltzmann's constant, $\theta_D$ is the Debye temperature, and $x$ is the phase parameter ($\hbar\omega/k_B$). The combination of one Einstein mode and two Debye modes fits the experimental data well, yielding oscillator terms consistent with the total number of atoms per formula unit. The total oscillator strengths are 9.8(5) and 8.5(6) for the Rb and Cs compounds, respectively (Table S6), slightly lower than the expected value of 12, which corresponds to the total number of atoms per formula unit in ANd(SO$_4$)$_2$. This discrepancy can be attributed to the collective behavior of atoms, as the strong covalent bonds between sulfur and oxygen cause them to behave more like a tetrahedral SO$_4$ group rather than as individual atoms.[37-39]

The field-dependent heat capacity (Figure 4c, d) shows an anomaly at low temperatures. This anomaly can be attributed to the nuclear quadrupole effect of Nd$^{3+}$,[40-43] as no magnetic transition was observed in the magnetization data (Figure S1, 2). To quantify the Schottky gap, the low-temperature phonon and a two-level Schottky model were fitted using equation 4:

$$\frac{C_p}{T} = (aT + bT^c) + \frac{R \times (\Delta/T)^2 \times \frac{g_0}{g_1} e^{\Delta/T}}{\left(1 + \frac{g_0}{g_1} e^{\Delta/T}\right)^2} \quad (4)$$

where $a$, $b$, and $c$ are constants describing the low-temperature phonon data, $R$ is the gas constant, $\Delta$ is the Schottky gap between two energy levels, and $g_0$ and $g_1$ represent the degeneracies of the ground and excited states, respectively. This model was selected because the data primarily captures the tail of the Schottky anomaly, where the energy levels are already partially populated in the measured temperature range. Moreover, the two-level Schottky model closely approximates the behavior expected from a full hyperfine coupling model. According to the fitted results (Table S7), the Schottky gap from the $C_p/T$ vs. $T$ curve is proportional to $\sqrt{(\langle\mu_s\rangle)^2 + \boldsymbol{B}^2}$ (Figure 4c-d inset), where the ⟨$\mu_s$⟩ represents the internal magnetic field from nuclear spin and $\boldsymbol{B}$ is the applied external field. The linear relationship suggests that ⟨$\mu_s$⟩ ≈ 0, consistent with the fact that approximately 80% of naturally occurring Nd isotopes possess zero nuclear spin.

The excitation spectra were collected at an emission wavelength of 392 nm for ANd(SO$_4$)$_2$ at 78 K and 300 K (Figure 5a, S3a, d). Six excitation peaks are observed between 240 and 375 nm (Figure S3a), attributable to the transitions from $^4I_{9/2}$ to $^3P$, $^2P$, $^2S$, $^2K$, $^2L$, and strong charge transfer from $4f^3$ to $4f^25d^1$. The characteristic emission of Nd$^{3+}$, $^4F_{3/2} \rightarrow {^4I_J}$, typically occurs in the 800-1400 nm range.[44-47] Emissions are rarely observed in the visible region. This can be attributed to 4$f$–4$f$ transitions being parity-forbidden and easily quenched by nonradiative processes such as cross-relaxation.[48, 49] In this process, excited ions exchange energy through Coulombic or exchange interactions, assisted by lattice vibrations, more efficiently than by photon emission or absorption, due to the small energy gaps between excited states (Figure 5a). This phenomenon commonly occurs in doped systems,[50, 51] particularly at high dopant concentrations. In contrast, in ANd(SO$_4$)$_2$, the Nd$^{3+}$ ions are relatively well-separated (> 5 Å) and experience a well-defined, homogeneous crystal field, leading to more clearly resolved energy levels (Figure 5b-c). This resolution enables the study of the otherwise broad emissions in the visible region.[44, 52] The emission intensity of both compounds shows a strong temperature dependence, decreasing with increasing temperature. As temperature increases, enhanced lattice vibrations (phonons) facilitate cross-relaxation, thereby quenching radiative emissions, consistent with the proposed mechanism. Meanwhile, the states are more clearly resolved in the Rb compound as sharp peaks in the emission spectra, whereas the Cs version shows broader peaks with shoulders. Since both compounds are isostructural, the crystal field splitting follows the same pattern (Figure 1), and the differences in spectra arise from the bonding difference due to the distinct A-site cations. Cs atoms, having higher electron-sharing tendencies, increase the electron density of oxygen atoms (Table S4). Consequently, improved interactions between Nd$^{3+}$ and SO$_4^{2-}$ groups occur in the Cs compound, making the Nd$^{3+}$ states in CsNd(SO$_4$)$_2$ more diffuse than those in RbNd(SO$_4$)$_2$ owing to better overlap with the ligand states.

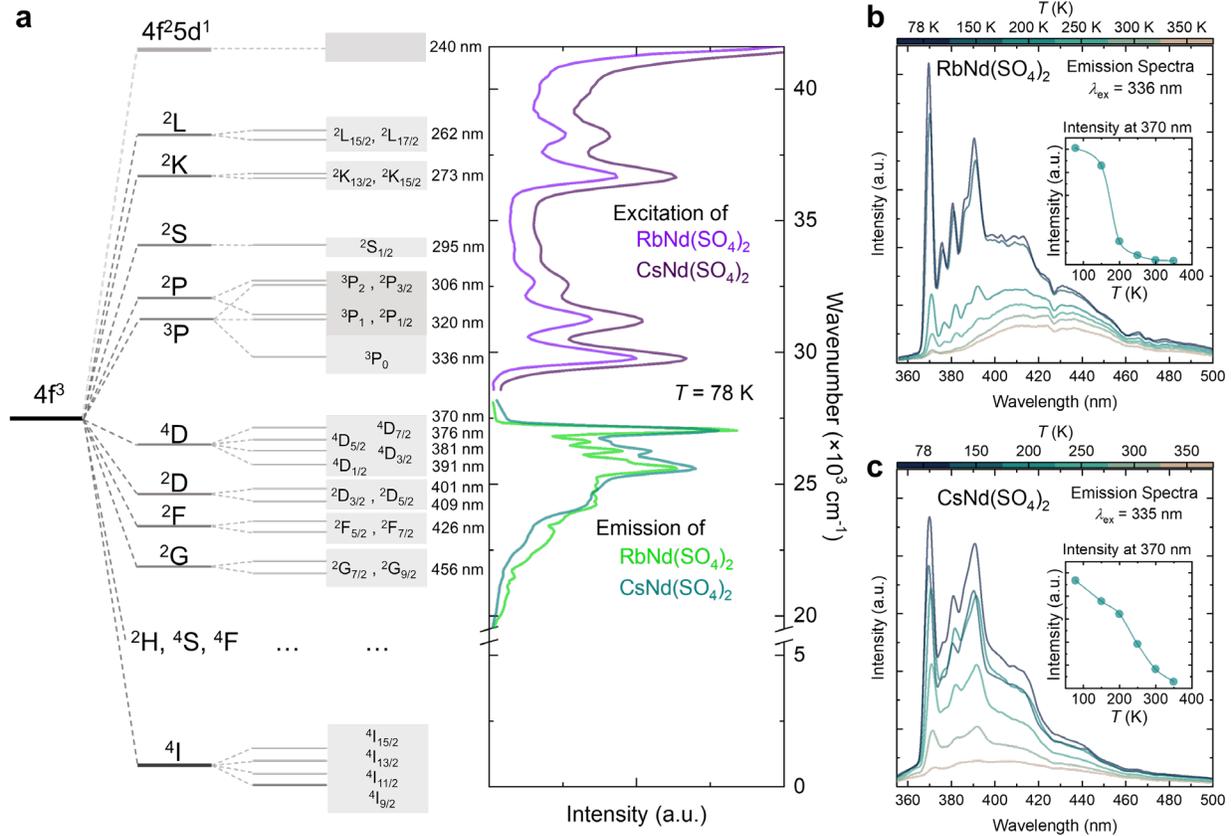

**Figure 5.** (a) Energy splitting diagram of Nd$^{3+}$ and excitation, emission spectra of RbNd(SO$_4$)$_2$ and CsNd(SO$_4$)$_2$ (c, d). Temperature dependent emission spectra with the inset showing the intensity of emission at 370 nm for ANd(SO$_4$)$_2$.

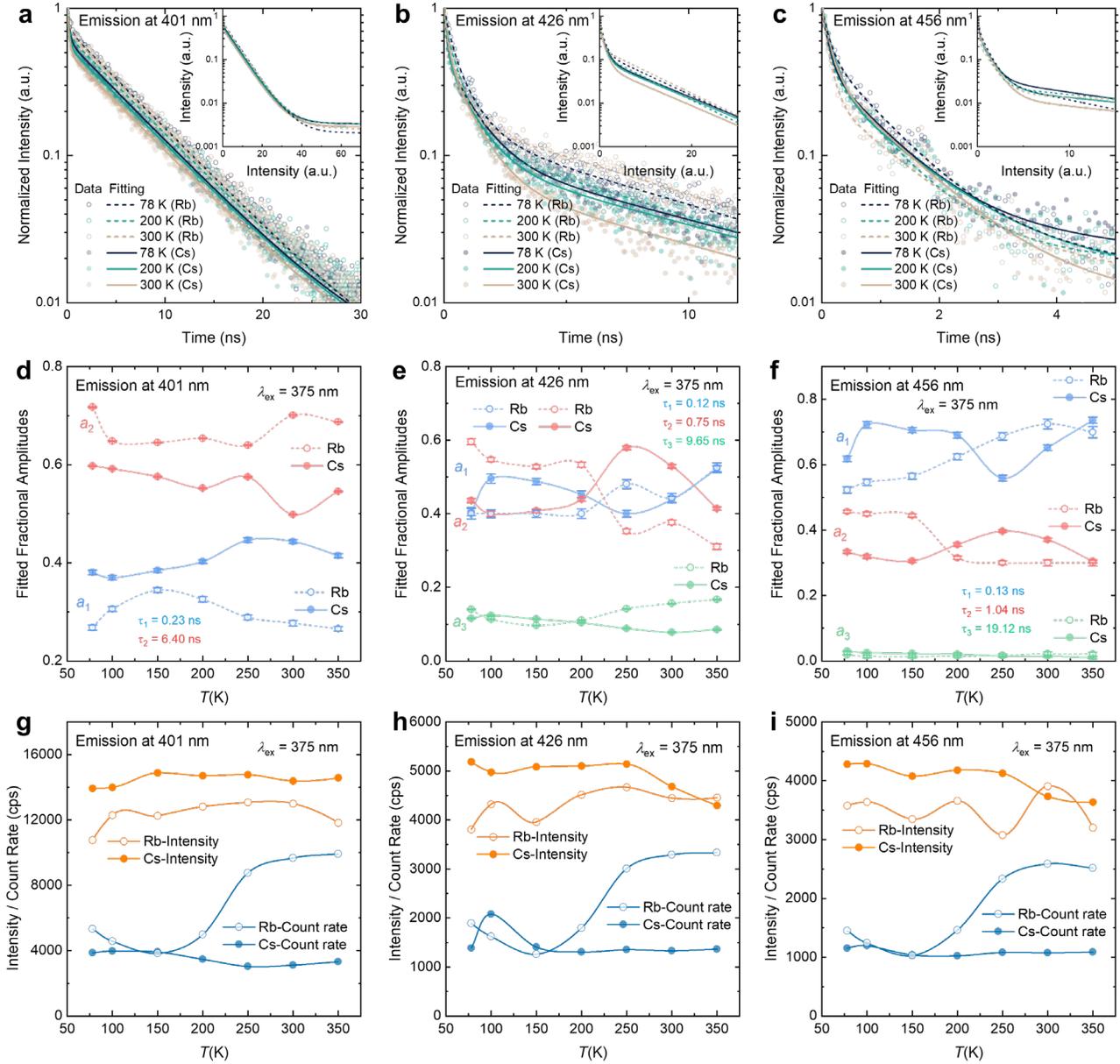

**Figure 6**. (a-c) Temperature dependent decay traces and fittings at (a) 401 nm, 426 nm, and 456 nm for ANd(SO$_4$)$_2$. (d-f) Fitted fractional amplitudes ($a_1$, $a_2$, $a_3$) as a function of temperature different emissions. (g-i) Temperature-dependent emission intensity (steady-state detector) and photon count rate (TRPL).

To further understand the temperature-dependent emission behavior and the role of phonons, time-resolved photoluminescence (TRPL) measurements were conducted on both samples at 78K ≤ T ≤ 300 K under 375 nm excitation (Figure 6). As shown in Figure 6a-c, the emission at 401 nm behaves differently from those at 426 and 456 nm. Therefore, the decay trace of the 401 nm emission was fitted using a bi-exponential model (Equation (5)), while a tri-exponential model (Equation (6)) was applied for the 426 and 456 nm emissions (Figure S4): [53, 54]

$$\frac{I}{I_0} = a_1 \cdot e^{(-t/\tau_1)} + a_2 \cdot e^{(-t/\tau_2)} \quad (5)$$

$$\frac{I}{I_0} = a_1 \cdot e^{(-t/\tau_1)} + a_2 \cdot e^{(-t/\tau_2)} + a_3 \cdot e^{(-t/\tau_3)} \quad (6)$$

where $I$ is the measured intensity, $I_0$ is the initial intensity, $a_1$, $a_2$, $a_3$ are the normalized fractional amplitudes for different exponential components, and $\tau_1$, $\tau_2$, $\tau_3$ are the corresponding lifetimes. Shorter lifetime components ($a_1$, $a_2$, $\tau_1$, $\tau_2$) reflect fast energy transfer and quenching via cross-relaxation or are impacted by nonradiative decays, while the longest lifetime component ($a_3$, $\tau_3$) corresponds to recombination from thermally isolated excited states (likely radiative). Although the broad emissions present some challenges in extracting lifetimes, the decay fitting provides valuable insights. Given the instrument resolution of 54.87 ps ($5.49 \times 10^{-2}$ ns per channel), all the extracted decay components are well within the detectable time window, indicating that the observed quenching arises from phonon-assisted nonradiative relaxation rather than ultrafast processes beyond the detection limit.

The emission associated with $^2D_{5/2, 3/2} \rightarrow {}^4I_{9/2}$ at 401 nm reveals fast exponential components ($\tau_1$, $\tau_2$) attributable to rapid cross-relaxation. While $a_{1Cs}$—the exponential component corresponding to $\tau_1$ of the Cs material—weighs more than $a_{1Rb}$, $a_{2Rb}$ is greater than $a_{2Cs}$.[44, 55] For the emissions associated with $^2F_{7/2, 5/2} \rightarrow {}^4I_{9/2}$ and $^2G_{9/2, 7/2} \rightarrow {}^4I_{9/2}$ at 426 and 456 nm, respectively, there is a long-lived component ($a_3$, $\tau_3$) observed, originating from additional relaxations from $^2D$ states. Notably, the fitted fractional amplitudes for the slow component ($a_3$) decrease as temperature increases, and $a_3$ is smaller in the Cs compound compared to Rb at high temperature, indicating a shorter lifetime for Cs at higher temperatures. To further probe the temperature-dependent emission dynamics, two complementary data sets were collected at 401, 426, and 456 nm using the same pulsed NanoLED excitation source ($\lambda = 375$ nm, 1 MHz repetition rate): one using a time-resolved detector to obtain TRPL photon count rates, and the other using a steady-state detector to measure emission intensity. Although both measurements used pulsed excitation, the steady-state detector integrates the signal over a fixed acquisition window without time resolution, providing a measure of temperature-dependent emission amplitude. In contrast, the time-resolved detector records the full decay profile after each pulse, allowing extraction of both lifetime information and the photon count rate, which reflects the number of detected photons per unit time and is sensitive to changes in both emission amplitude and lifetime. The results are shown in Figure 6g–i. For both compounds, the steady-state PL intensities exhibit a gradual decrease with increasing temperature, consistent with conventional thermal quenching. In contrast, the TRPL count rates reveal more complex, compound-dependent behaviors. The Cs compound shows an increase in count rate up to ~100 K, followed by a decline at higher temperatures. The Rb compound, however, displays an initial decrease between 78–150 K, a recovery between 150–250 K, and a final decrease above 250 K. These non-monotonic TRPL trends suggest that additional factors beyond thermal quenching influence the excited-state dynamics and point to the role of phonons. The quenching of emission intensity with increasing temperature can be attributed to phonon-assisted nonradiative relaxation, consistent with the lattice contributions revealed by the heat capacity (Figure 4a, b). The Einstein and Debye(1) modes primarily contribute below 78 K and are mostly saturated over the temperature range of the PL measurements, thus having minimal influence on the observed emission behavior. Notably, the different trends in photon count rates between the two compounds can be rationalized by their respective Debye(2) behaviors extracted from the phonon model. The Debye(2) mode in $CsNd(SO_4)_2$ ($\theta_{D(2)} = 667$ K) begins saturating near 120 K, whereas in $RbNd(SO_4)_2$, the higher $\theta_{D(2)}$ value of 964 K implies that phonon population continues to increase up to 250 K. This extended phonon activity in the Rb compound may enable thermally activated repopulation of radiative states or facilitate delayed emission recovery, consistent with the observed rise in count rate above 150 K. Generally, for steady-state PL measurements, both compounds show a slight decrease with increasing temperature (Figure 6g-i). In contrast, the TRPL count rate for $CsNd(SO_4)_2$ shows a local maximum near 100 K, followed by a gradual decline that mirrors the steady-state trend at higher temperatures. $RbNd(SO_4)_2$ exhibits a distinct dip in TRPL count rate between 78 and 150 K, followed by a notable increase from 150 to 250 K, before transitioning to a similar weakly decreasing trend as seen in steady-state PL. These contrasting TRPL behaviors indicate phonon-dependent modulation of excited-state lifetimes. We interpret these behaviors as signatures of phonon-assisted modulation of excited-state lifetime: initial phonon population reduces nonradiative losses by slowing deactivation, whereas full phonon saturation at higher temperatures facilitates energy dissipation and lifetime shortening. This analysis reveals a key connection between lattice vibrations and optical deactivation pathways, emphasizing the utility of TRPL as a sensitive probe of phonon–exciton coupling in low-symmetry lanthanide systems. This observation aligns with the proposed notion that the interaction between $Nd^{3+}$ and $SO_4^{2-}$ is more effective in the Cs compound, consistent with the integrated crystal orbital bond index (ICOBI) results.

To gain deeper insights into the electronic structure of $ANd(SO_4)_2$ (A = Rb or Cs), full-potential spin-polarized DFT calculations were performed using WIEN2k (Figure 7).[56] The results reveal several common characteristics of $ANd(SO_4)_2$. Bands near the Fermi level ($E_F$) originating from Cs, S, and O are relatively diffused, indicating strong overlap between the Cs–O and S–O bonds. The Nd-$f$ orbitals are contracted, consistent with the spherical nature of the $f$ orbitals. The spins of the Nd-$f$ states are polarized, subsequently polarizing the O-$p$, S-$p$, and A-$s$ (A = Rb or Cs) states. The bands in the Cs compound are more diffuse than those in the Rb compound, supporting the observation of greater covalency in the Cs material.

The density of states (DOS) provides valuable insights into the contributions of different states but does not include the phase information of orbitals involved in wavefunction overlaps (either constructive or destructive interference). To overcome this limitation, the LOBSTER program[57, 58] enables the reconstruction of wavefunctions to reveal the critical phase information in $ANd(SO_4)_2$. Consequently, pseudopotential DFT calculations were carried out using the projector-augmented wave (PAW) method in Quantum Espresso software, followed by projection into a linear combination of atomic orbitals (LCAO) using LOBSTER. The projected crystal orbital Hamilton population (-pCOHP)[59] curves indicate nonbonding interactions at the Fermi level ($E_F$) (Figure S5), corroborating the DOS data and confirming the insulating nature of these compounds. The spin-polarized -pCOHP curves for Rb-O and Cs-O show similar trends but differ in magnitude, suggesting stronger interactions between Cs and O near the $E_F$. Integrated COHP (-ICOHP) values, which describe the overall strength of bonding, are large and positive (Figure 8a, S6), indicating robust bonding within ANd(SO$_4$)$_2$ compounds. The bonding strength is slightly lower in the Cs-based compound, likely due to a larger crystal unit cell and longer bond lengths. To further understand the nature of these bonds and validate our assumptions about covalency differences, we calculated the Integrated Crystal Orbital Bond Index (ICOBI), which measures the number of electrons shared during bond formation (indicative of covalency).[60, 61] Our findings (Figure 8b) reveal that the A–O and Nd–O bonds exhibit low ICOBI values, suggesting predominantly ionic bonding, whereas the S–O bond shows approximately one electron shared, indicating covalent bonding. Moreover, consistent with predictions based on the inductive effect, the Cs compound exhibits higher covalency than the Rb compound.

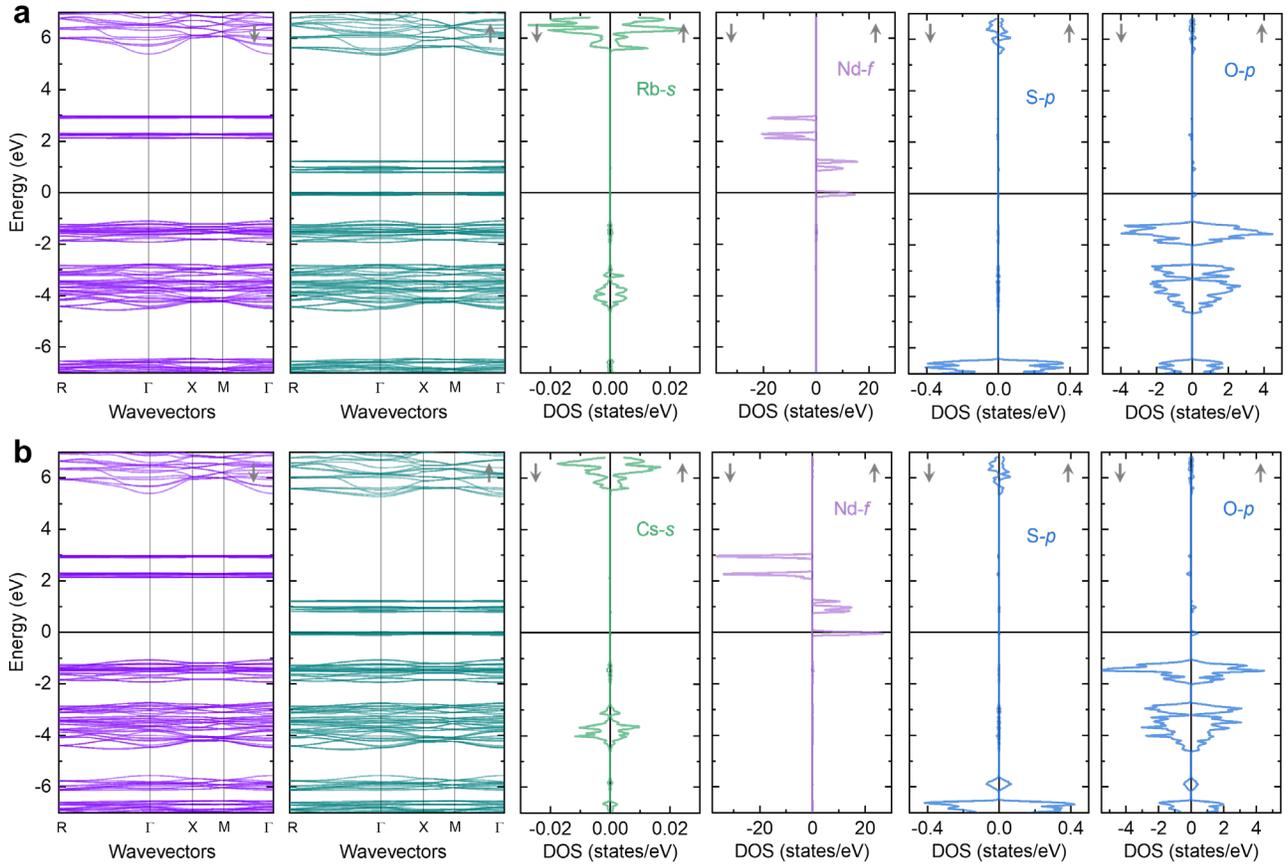

**Figure 7**. Spin-polarized band structure and spin polarized density of states (DOS) for (a) RbNd(SO$_4$)$_2$ and (b) RbNd(SO$_4$)$_2$.

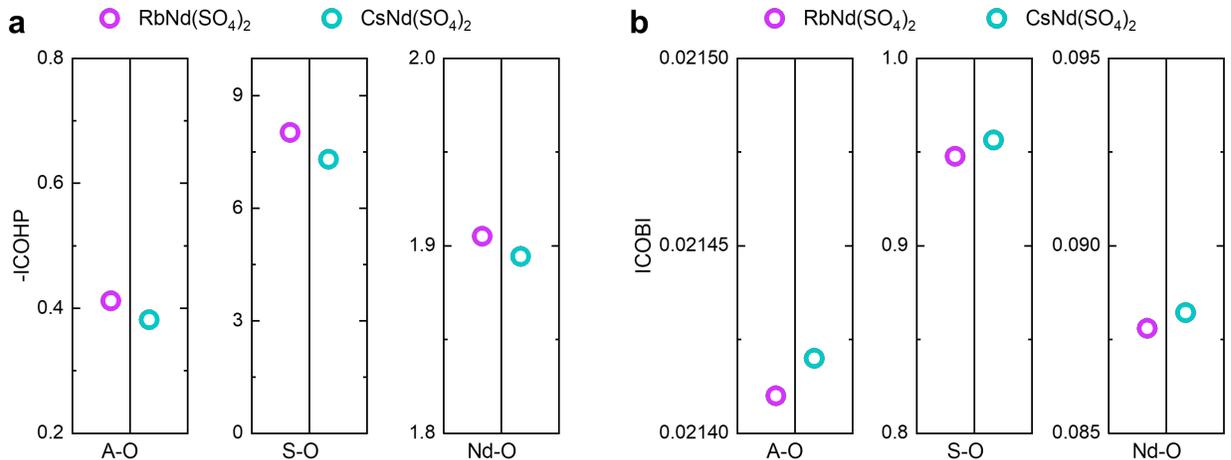

**Figure 8**. (a). Integrated crystal orbital Hamilton population (-ICOHP), showing the strength of the bonding. (b). Integrated crystal orbital bond index (ICOBI), showing the covalency of the bonding.

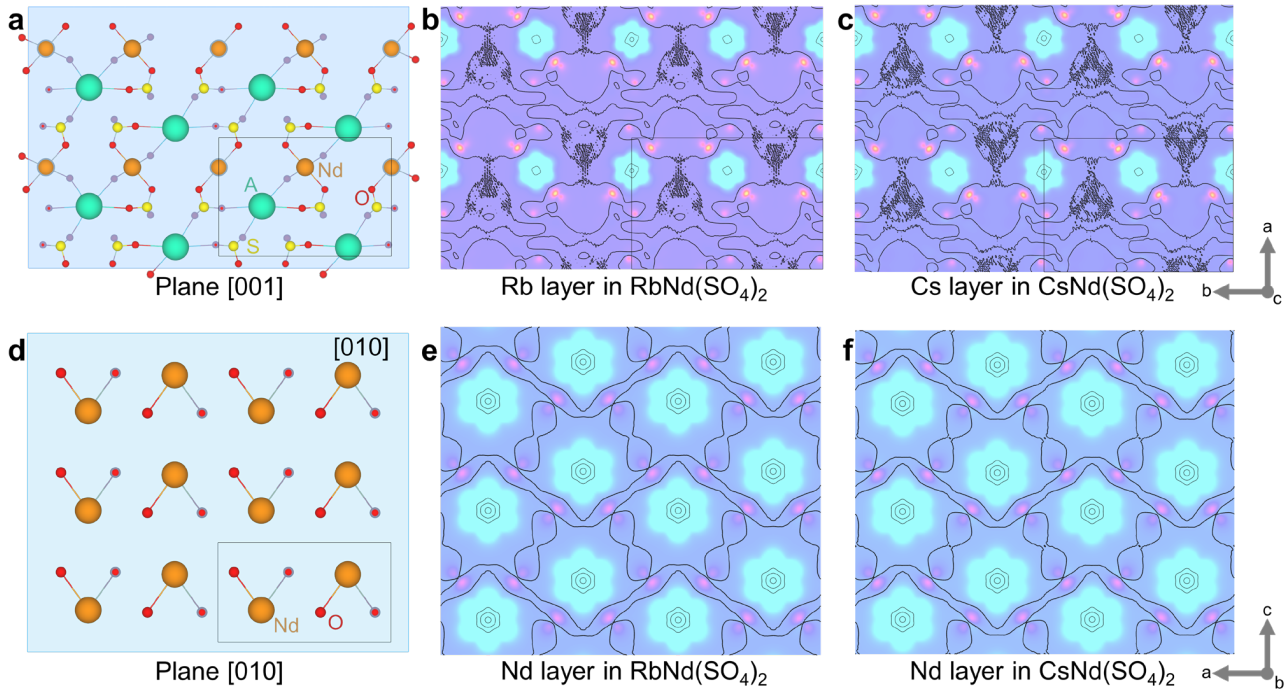

**Figure 9**. Spin density map on (a-c) [001] plane cutting through the A layer showing polarized Rb and Cs. (d-f) [010] plane on the distorted Nd triangular layer showing magnetic interactions within the layer.

The intralayer and interlayer exchange pathways were mapped out using the spin density map ($\rho_{up} - \rho_{down}$) and projected on selected lattice planes. Figure 9a-c highlights the spin polarization on the [001] plane (interlayer). The Nd-$f$ magnetic spins polarize the spin density of the O-$p$ states, which then polarize the A-site cations. The spin polarization is stronger in the Cs compound compared to Rb, suggesting enhanced interlayer interactions in CsNd(SO$_4$)$_2$. Figure 9d-f illustrates the spin polarization on the [010] plane (intralayer). The polarized Nd-$f$ spins clearly form a triangular lattice, and the strong magnetic interactions occur within the layers, as evidenced by the magnetization data.

## CONCLUSION

ANd(SO$_4$)$_2$ (A = Rb, Cs) compounds exhibit tunable optical, magnetic, and electronic properties arising from their distorted triangular lattice structure and the inductive effect. Magnetization measurements indicate sizable AFM interactions, yet no long-range magnetic ordering is observed down to 1.8 K. Temperature-dependent photoluminescence studies reveal that the lattice dynamics and bonding of the Rb and Cs compounds strongly influence cross-relaxation processes. TRPL measurements further demonstrate distinct recombination dynamics, with faster nonradiative relaxation observed in the Cs compound due to enhanced overlap between Nd and ligands. Heat capacity proves no magnetic ordering while illuminating insight into the role of phonons in emission relaxation. The inductive effect, introduced by the A-site, enables the modification of magnetic, electronic, and optical properties while preserving the uniqueness of triangular spin lattices. DFT calculations support the experimental results, revealing enhanced covalency and bonding interactions in the Cs compound compared to the Rb analog. These findings contribute to a deep understanding of how subtle chemical modifications can influence many-body electronic states, phonons, and emission lifetime in low-dimensional frustrated magnets. Further studies in this system may uncover new magnetic phenomena and offer pathways for uniting multiple functions in quantum materials.

## EXPERIMENTAL SECTION

**SYNTHESIS.** The synthesis and characterization of TAS precursor used as the sulfate source have been reported by us.[62, 63] The ANd(SO$_4$)$_2$ (A = Rb, Cs) crystals were prepared by hydrothermal reaction. 2 mmol of ACl (A = Rb, Cs) and 2 mmol of Nd(NO$_3$)$_3$·6H$_2$O were dissolved in 10 mL of 4M HNO$_3$ in a 23 mL PTFE-lined autoclave, and 2 mmol of TAS was added. The sealed autoclaves were heated to 200℃ in 2 hours, kept at 200℃ for 60 hours, and cooled to room temperature at the rate of 5 ℃ per hour. Single-phase plate-like pale pink transparent crystals were obtained in the autoclave.

**Single-crystal X-ray diffraction (XRD).** Single crystal diffraction experiments were performed on ANd(SO$_4$)$_2$ (A = Rb, Cs) using a Bruker D8 Venture diffractometer with Mo Kα radiation (λ = 0.71073 Å) and a Photon 100 detector at T = 100 K. Data processing (SAINT) and scaling (SADABS) were performed using the Apex3 software. The structure was solved by the intrinsic phasing 2 method (SHELXT) and refined by full-matrix least-squares techniques on F2 (SHELXL) using the SHELXTL software suite. All atoms were refined anisotropically.

**Magnetization and Specific Heat.** DC magnetization measurements on ANd(SO$_4$)$_2$ (A = Rb, Cs) powder were performed with the vibrating sample magnetometer (VSM) option of the Quantum Design Physical Properties Measurement System (PPMS). Heat capacity was measured using the PPMS, employing the semiadiabatic pulse technique from $T$ = 2 to 300 K.

**Temperature-dependent photoluminescent.** The temperature-dependent photoluminescence (PL) experiments were conducted using a HORIBA Nanolog modular spectrofluorometer system.[64] This setup included a Xenon lamp (HORIBA FL-1039), an excitation monochromator, a sample-compartment module, an emission spectrometer, and a CCD detector (HORIBA Symphony II). TRPL spectra were captured using a HORIBA FluoroLog FL3-22 spectrofluorometer, which was equipped with a picosecond photon detection module (TBX-05) and a 375 nm laser source (NanoLED N-375L), managed by a single photon counting controller (FluoroHub). The photoluminescence spectra were corrected for the spectral response of the monochromator and detector. For TRPL, photon arrival times were collected using time-correlated single-photon counting (TCSPC), with count rates maintained below 1% of the laser repetition rate to avoid pile-up. Data were normalized prior to exponential fitting.

For sample preparation, a suspension of $ANd(SO_4)_2$ in methanol was drop-cast onto quartz substrates and dried at 50°C for 5 minutes. The substrates were then placed inside an optical cryostat (JANIS, ST-100 equipped with a Turbolab 350), illuminated by the Xenon lamp and laser. The temperature inside the cryostat was meticulously controlled using a Lake Shore temperature controller (Model 325) and cooled with liquid nitrogen to achieve the desired experimental conditions.

**Density functional theory calculation.** Spin-polarized electronic structure calculations were performed using a full-potential linearized augmented plane wave method by the WIEN2k code[56]. The exchange and correlation energies were treated within the DFT using the Perdew–Burke–Ernzerhof generalized gradient approximation[65]. The muffin-tin radius values 2.30, 2.30, 2.30, 1.31 and 1.31 au were used for Rb, Cs, Nd, S and O, respectively. The self-consistencies were carried out using 3 × 2 × 5 mesh, in the irreducible Brillouin zone. The density of states, crystal orbital Hamiltonian population (COHP) and crystal orbital bond index were calculated using the Quantum Espresso software package[66] with the Generalized Gradient Approximation (GGA+U)[67] of the exchange-correlation potential with the PBEsol parametrization[68]. Projector-augmented wave (PAW) potentials for Rb, Cs, Nd, S and O were taken from the PSlibrary v.1.0.0 set[69]. The k-mesh utilized is the same as that in WIEN2k calculations. Kinetic energy cutoff for charge density and wavefunctions was set to 34 eV and 323 eV. The projected crystal orbital Hamilton populations (-pCOHP), and crystal orbital bond index (COBI) with their integrated value (-ICOHP, and ICOBI) up to $E_F$ are calculated with LOBSTER software[57, 58] to extract bonding information.

## ASSOCIATED CONTENT

**Supporting Information**. Additional information on the crystal structure, optical and magnetic properties, and DFT calculation of $RbNd(SO_4)_2$ and $CsNd(SO_4)_2$; single crystal data and calculated atomic charge, excitation and emission spectra, TRPL, magnetic susceptivity data under applied field; magnetic properties of $Nd^{3+}$ in triangular lattice; phonon and Schottky effect fitting; crystal orbital Hamilton population of $RbNd(SO_4)_2$ and $CsNd(SO_4)_2$. This material is available free of charge via the Internet at http://pubs.acs.org.

## AUTHOR INFORMATION

Corresponding Author

* thao@clemson.edu

Author Contributions

The manuscript was written through contributions of all authors. All authors have given approval to the final version of the manuscript.

## ACKNOWLEDGMENT

The work was supported by the Arnold and Mabel Beckman Foundation through a 2023 BYI award to TTT. We also thank the NSF Awards NSF-DMR-2338014 and NSF-OIA-2227933 and the Camille Henry Dreyfus Foundation for the support. HS and TTT acknowledge funds from the Clemson Major Research Instrumentation (CU-MRI) initiative.

## ABBREVIATIONS

DFT, density functional theory; TRPL, time-resolved photoluminescence; SOC, spin-orbital coupling

## REFERENCES

(1) Ramirez, A. P. Strongly geometrically frustrated magnets. *Annual Review of Materials Science* **1994**, *24* (1), 453-480.
(2) Balents, L. Spin liquids in frustrated magnets. *nature* **2010**, *464* (7286), 199-208.
(3) Chamorro, J. R.; McQueen, T. M.; Tran, T. T. Chemistry of quantum spin liquids. *Chemical Reviews* **2020**, *121* (5), 2898-2934.
(4) Tokura, Y.; Kanazawa, N. Magnetic skyrmion materials. *Chemical Reviews* **2020**, *121* (5), 2857-2897.
(5) Jackson, A.; Jackson, A. D.; Pasquier, V. The skyrmion-skyrmion interaction. *Nuclear Physics A* **1985**, *432* (3), 567-609.
(6) Jiang, W.; Zhang, X.; Yu, G.; Zhang, W.; Wang, X.; Benjamin Jungfleisch, M.; Pearson, J. E.; Cheng, X.; Heinonen, O.; Wang, K. L. Direct observation of the skyrmion Hall effect. *Nature Physics* **2017**, *13* (2), 162-169.
(7) Mila, F.; Poilblanc, D.; Bruder, C. Spin dynamics in a frustrated magnet with short-range order. *Physical Review B* **1991**, *43* (10), 7891.
(8) Ye, F.; Fernandez-Baca, J. A.; Fishman, R. S.; Ren, Y.; Kang, H. J.; Qiu, Y.; Kimura, T. Magnetic interactions in the geometrically frustrated triangular lattice antiferromagnet $CuFeO_2$. *Physical review letters* **2007**, *99* (15), 157201.
(9) Kumar, S.; van den Brink, J. Frustration-induced insulating chiral spin state in itinerant triangular-lattice magnets. *Physical review letters* **2010**, *105* (21), 216405.
(10) Karunadasa, H.; Huang, Q.; Ueland, B. G.; Lynn, J. W.; Schiffer, P.; Regan, K. A.; Cava, R. J. Honeycombs of

triangles and magnetic frustration in SrL$_2$O$_4$ (L= Gd, Dy, Ho, Er, Tm, and Yb). *Physical Review B* **2005**, *71* (14), 144414.
(11) Sanjeewa, L. D.; Xing, J.; Taddei, K. M.; Sefat, A. S. Synthesis, crystal structure and magnetic properties of KLnSe$_2$ (Ln= La, Ce, Pr, Nd) structures: A family of 2D triangular lattice frustrated magnets. *Journal of Solid State Chemistry* **2022**, *308*, 122917.
(12) Oyeka, E. E.; Winiarski, M. J.; Błachowski, A.; Taddei, K. M.; Scheie, A.; Tran, T. T. Potential skyrmion host Fe(IO$_3$)$_3$: Connecting stereoactive lone-pair electron effects to the Dzyaloshinskii-Moriya interaction. *Chemistry of Materials* **2021**, *33* (12), 4661-4671.
(13) Huai, X.; Acheampong, E.; Delles, E.; Winiarski, M. J.; Sorolla, M.; Nassar, L.; Liang, M.; Ramette, C.; Ji, H.; Scheie, A. Noncentrosymmetric Triangular Magnet CaMnTeO$_6$: Strong Quantum Fluctuations and Role of s$^0$ versus s$^2$ Electronic States in Competing Exchange Interactions. *Advanced Materials* **2024**, 2313763.
(14) Qureshi, N.; Wildes, A. R.; Ritter, C.; Fåk, B.; Riberolles, S. X. M.; Hatnean, M. C.; Petrenko, O. A. Magnetic structure and low-temperature properties of geometrically frustrated SrNd$_2$O$_4$. *Physical Review B* **2021**, *103* (13), 134433.
(15) Chen, P.; Holinsworth, B. S.; O'Neal, K. R.; Luo, X.; Topping, C. V.; Cheong, S. W.; Singleton, J.; Choi, E. S.; Musfeldt, J. L. Frustration and Glasslike Character in RIn$_{1-x}$Mn$_x$O$_3$ (R= Tb, Dy, Gd). *Inorganic chemistry* **2018**, *57* (20), 12501-12508.
(16) Ortiz, B. R.; Bordelon, M. M.; Bhattacharyya, P.; Pokharel, G.; Sarte, P. M.; Posthuma, L.; Petersen, T.; Eldeeb, M. S.; Granroth, G. E.; Dela Cruz, C. R. Electronic and structural properties of RbCe$X_2$ ($X_2$: O$_2$, S$_2$, SeS, Se$_2$, TeSe, Te$_2$). *Physical Review Materials* **2022**, *6* (8), 084402.
(17) Etourneau, J.; Portier, J.; Ménil, F. The role of the inductive effect in solid state chemistry: how the chemist can use it to modify both the structural and the physical properties of the materials. *Journal of alloys and compounds* **1992**, *188*, 1-7.
(18) Sun, W.; Bartel, C. J.; Arca, E.; Bauers, S. R.; Matthews, B.; Orvañanos, B.; Chen, B.-R.; Toney, M. F.; Schelhas, L. T.; Tumas, W. A map of the inorganic ternary metal nitrides. *Nature materials* **2019**, *18* (7), 732-739.
(19) Mei, D.; Cao, W.; Wang, N.; Jiang, X.; Zhao, J.; Wang, W.; Dang, J.; Zhang, S.; Wu, Y.; Rao, P. Breaking through the "3.0 eV wall" of energy band gap in mid-infrared nonlinear optical rare earth chalcogenides by charge-transfer engineering. *Materials Horizons* **2021**, *8* (8), 2330-2334.
(20) Harada, J. K.; Charles, N.; Poeppelmeier, K. R.; Rondinelli, J. M. Heteroanionic materials by design: Progress toward targeted properties. *Advanced Materials* **2019**, *31* (19), 1805295.
(21) Culver, S. P.; Squires, A. G.; Minafra, N.; Armstrong, C. W. F.; Krauskopf, T.; Böcher, F.; Li, C.; Morgan, B. J.; Zeier, W. G. Evidence for a solid-electrolyte inductive effect in the superionic conductor Li$_{10}$Ge$_{1-x}$Sn$_x$P$_2$S$_{12}$. *Journal of the American Chemical Society* **2020**, *142* (50), 21210-21219.
(22) Kent, G. T.; Morgan, E.; Albanese, K. R.; Kallistova, A.; Brumberg, A.; Kautzsch, L.; Wu, G.; Vishnoi, P.; Seshadri, R.; Cheetham, A. K. Elusive double perovskite iodides: structural, optical, and magnetic properties. *Angewandte Chemie International Edition* **2023**, *62* (32), e202306000.
(23) Xie, M.; Zhuo, W.; Cai, Y.; Zhang, Z.; Zhang, Q. Rare-Earth Chalcogenides: An Inspiring Playground for Exploring Frustrated Magnetism. *Chinese Physics Letters* **2024**, *41* (11), 117505.
(24) Qureshi, N.; Wildes, A. R.; Ritter, C.; Fåk, B.; Riberolles, S. X. M.; Hatnean, M. C.; Petrenko, O. A. Magnetic structure and low-temperature properties of geometrically frustrated SrNd$_2$O$_4$. *Physical Review B* **2021**, *103* (13), 134433.
(25) Sawicki, B.; Tomaszewicz, E.; Groń, T.; Oboz, M.; Kusz, J.; Berkowski, M. Magnetic and electrical characteristics of Nd$^{3+}$-doped lead molybdato-tungstate single crystals. *Materials* **2023**, *16* (2), 620.
(26) Misra, S. K.; Isber, S. EPR of the Kramers ions Er$^{3+}$, Nd$^{3+}$, Yb$^{3+}$ and Ce$^{3+}$ in Y(NO$_3$)$_3$·6H$_2$O and Y$_2$(SO$_4$)$_3$·8H$_2$O single crystals:: Study of hyperfine transitions. *Physica B: Condensed Matter* **1998**, *253* (1-2), 111-122.
(27) Malakhovskii, A. V.; Edelman, I. S.; Radzyner, Y.; Yeshurun, Y.; Potseluyko, A. M.; Zarubina, T. V.; Zamkov, A. V.; Zaitzev, A. I. Magnetic and magneto-optical properties of oxide glasses containing Pr$^{3+}$, Dy$^{3+}$ and Nd$^{3+}$ ions. *Journal of magnetism and magnetic materials* **2003**, *263* (1-2), 161-172.
(28) Kumar, K. U.; Prathyusha, V. A.; Babu, P.; Jayasankar, C. K.; Joshi, A. S.; Speghini, A.; Bettinelli, M. Fluorescence properties of Nd$^{3+}$-doped tellurite glasses. *Spectrochimica Acta Part A: Molecular and Biomolecular Spectroscopy* **2007**, *67* (3-4), 702-708.
(29) Kumar, K. V.; Kumar, A. S. Spectroscopic properties of Nd$^{3+}$ doped borate glasses. *Optical Materials* **2012**, *35* (1), 12-17.
(30) Lubinskii, N. N.; Bashkirov, L. A.; Galyas, A. I.; Shevchenko, S. V.; Petrov, G. S.; Sirota, I. M. Magnetic susceptibility and effective magnetic moment of the Nd$^{3+}$ and Co$^{3+}$ ions in NdCo$_{1-x}$Ga$_x$O$_3$. *Inorganic Materials* **2008**, *44* (9), 1015-1021.
(31) Strobel, S.; Choudhury, A.; Dorhout, P. K.; Lipp, C.; Schleid, T. Rare-earth metal (III) oxide selenides M$_4$O$_4$Se[Se$_2$](M= La, Ce, Pr, Nd, Sm) with discrete diselenide units: crystal structures, magnetic frustration, and other properties. *Inorganic chemistry* **2008**, *47* (11), 4936-4944.
(32) Chamorro, J. R.; Jackson, A. R.; Watkins, A. K.; Seshadri, R.; Wilson, S. D. Magnetic order in the $S_{eff}$= 1/2 triangular-lattice compound NdCd$_3$P$_3$. *Physical Review Materials* **2023**, *7* (9), 094402.
(33) Sun, W.; Huang, Y.-X.; Pan, Y.; Mi, J.-X. Strong spin frustration and negative magnetization in LnCu$_3$(OH)$_6$Cl$_3$ (Ln= Nd and Sm) with triangular lattices: the effects of lanthanides. *Dalton Transactions* **2017**, *46* (29), 9535-9541.

(34) Song, F.; Liu, A.; Chen, Q.; Zhou, J.; Li, J.; Tong, W.; Wang, S.; Wang, Y.; Lu, H.; Yuan, S. $Ba_6RE_2Ti_4O_{17}$ (RE= Nd, Sm, Gd, Dy–Yb): A Family of Rare-Earth-Based Layered Triangular Lattice Magnets. *Inorganic Chemistry* **2024**, *63* (13), 5831-5841.
(35) Ashtar, M.; Gao, Y. X.; Wang, C. L.; Qiu, Y.; Tong, W.; Zou, Y. M.; Zhang, X. W.; Marwat, M. A.; Yuan, S. L.; Tian, Z. M. Synthesis, structure and magnetic properties of rare-earth $REMgAl_{11}O_{19}$ (RE= Pr, Nd) compounds with two-dimensional triangular lattice. *Journal of Alloys and Compounds* **2019**, *802*, 146-151.
(36) Arh, T.; Sana, B.; Pregelj, M.; Khuntia, P.; Jagličić, Z.; Le, M. D.; Biswas, P. K.; Manuel, P.; Mangin-Thro, L.; Ozarowski, A. The Ising triangular-lattice antiferromagnet neodymium heptatantalate as a quantum spin liquid candidate. *Nature Materials* **2022**, *21* (4), 416-422.
(37) Fazzi, D.; Canesi, E. V.; Negri, F.; Bertarelli, C.; Castiglioni, C. Biradicaloid Character of Thiophene‐Based Heterophenoquinones: The Role of Electron‐Phonon Coupling. *ChemPhysChem* **2010**, *11* (17), 3685-3695.
(38) Degiorgi, L.; Alavi, B.; Mihály, G.; Grüner, G. Complete excitation spectrum of charge-density waves: Optical experiments on $K_{0.3}MoO_3$. *Physical Review B* **1991**, *44* (15), 7808.
(39) Chahal, J.; Rahbany, N.; El-Helou, Y.; Wu, K. T.; Bruyant, A.; Zgheib, C.; Kazan, M. Temperature dependence of the anisotropy of the infrared dielectric properties and phonon-plasmon coupling in n-doped 4H-SiC. *Journal of Physics and Chemistry of Solids* **2024**, *187*, 111861.
(40) Cheng, J. G.; Sui, Y.; Qian, Z. N.; Liu, Z. G.; Miao, J. P.; Huang, X. Q.; Lu, Z.; Li, Y.; Wang, X. J.; Su, W. H. Schottky-like anomaly in the low-temperature specific heat of single-crystal $NdMnO_3$. *Solid state communications* **2005**, *134* (6), 381-384.
(41) Bartolomé, J.; Palacios, E.; Kuz'Min, M. D.; Bartolomé, F.; Sosnowska, I.; Przeniosło, R.; Sonntag, R.; Lukina, M. M. Single-crystal neutron diffraction study of Nd magnetic ordering in $NdFeO_3$ at low temperature. *Physical Review B* **1997**, *55* (17), 11432.
(42) Gordon, J. E.; Fisher, R. A.; Jia, Y. X.; Phillips, N. E.; Reklis, S. F.; Wright, D. A.; Zettl, A. Specific heat of $Nd_{0.67}Sr_{0.33}MnO_3$. *Physical Review B* **1999**, *59* (1), 127.
(43) Brugger, T.; Schreiner, T.; Roth, G.; Adelmann, P.; Czjzek, G. Heavy-fermion-like excitations in metallic $Nd_{2-y}Ce_yCuO_4$. *Physical review letters* **1993**, *71* (15), 2481.
(44) Campbell, J. H.; Suratwala, T. I. Nd-doped phosphate glasses for high-energy/high-peak-power lasers. *Journal of non-crystalline solids* **2000**, *263*, 318-341.
(45) Zhao, C.; Zhang, L.; Hang, Y.; He, X.; Yin, J.; Hu, P.; Chen, G.; He, M.; Huang, H.; Zhu, Y. Optical spectroscopy of $Nd^{3+}$ in $LiLuF_4$ single crystals. *Journal of Physics D: Applied Physics* **2010**, *43* (49), 495403.
(46) Chen, X.; Luo, Z.; Jaque, D.; Romero, J. J.; Sole, J. G.; Huang, Y.; Jiang, A.; Tu, C. Comparison of optical spectra of $Nd^{3+}$ in $NdAl_3(BO_3)_4$ (NAB), Nd: $GdAl_3(BO_3)_4$ (NGAB) and Nd: $Gd_{0.2}Y_{0.8}Al_3(BO_3)_4$ (NGYAB) crystals. *Journal of Physics: Condensed Matter* **2001**, *13* (5), 1171.
(47) Wang, C.; Zhao, N.; Zhang, H.; Zhang, X.; Lin, X.; Liu, H.; Dang, F.; Zhang, W.; Sun, J.; Chen, P. Ultrawide UV to NIR Emission in Double Perovskite Nanocrystals via the Self-Trapping State Engineering Strategy. *ACS Sustainable Chemistry & Engineering* **2023**, *11* (40), 14659-14666.
(48) Reddy, C. M.; Raju, B. D. P.; Sushma, N. J.; Dhoble, N. S.; Dhoble, S. J. A review on optical and photoluminescence studies of $RE^{3+}$ (RE= Sm, Dy, Eu, Tb and Nd) ions doped LCZSFB glasses. *Renewable and Sustainable Energy Reviews* **2015**, *51*, 566-584.
(49) İlhan, M.; Keskin, İ. Ç.; Çatalgöl, Z.; Samur, R. NIR photoluminescence and radioluminescence characteristics of $Nd^{3+}$ doped $BaTa_2O_6$ phosphor. *International Journal of Applied Ceramic Technology* **2018**, *15* (6), 1594-1601.
(50) Jafarirad, S.; Salmasi, M.; Divband, B.; Sarabchi, M. Systematic study of $Nd^{3+}$ on structural properties of ZnO nanocomposite for biomedical applications; in-vitro biocompatibility, bioactivity, photoluminescence and antioxidant properties. *Journal of Rare Earths* **2019**, *37* (5), 508-514.
(51) Maia, L. J. Q.; Moura, A. L.; Jerez, V.; De Araujo, C. B. Structural properties and near infrared photoluminescence of $Nd^{3+}$ doped $YBO_3$ nanocrystals. *Optical Materials* **2019**, *95*, 109227.
(52) Falcao-Filho, E. L.; de Araujo, C. B.; Messaddeq, Y. Frequency upconversion involving triads and quartets of ions in a $Pr^{3+}/Nd^{3+}$ codoped fluoroindate glass. *Journal of applied physics* **2002**, *92* (6), 3065-3070.
(53) Holly Haggar, J. I.; Ghataora, S. S.; Trinito, V.; Bai, J.; Wang, T. Study of the luminescence decay of a semipolar green light-emitting diode for visible light communications by time-resolved electroluminescence. *ACS photonics* **2022**, *9* (7), 2378-2384.
(54) Péan, E. V.; Dimitrov, S.; De Castro, C. S.; Davies, M. L. Interpreting time-resolved photoluminescence of perovskite materials. *Physical Chemistry Chemical Physics* **2020**, *22* (48), 28345-28358.
(55) Liu, C. H.; Das, B. B.; Glassman, W. L. S.; Tang, G. C.; Yoo, K. M.; Zhu, H. R.; Akins, D. L.; Lubicz, S. S.; Cleary, J.; Prudente, R. Raman, fluorescence, and time-resolved light scattering as optical diagnostic techniques to separate diseased and normal biomedical media. *Journal of Photochemistry and Photobiology B: Biology* **1992**, *16* (2), 187-209.
(56) Schwarz, K.; Blaha, P.; Madsen, G. K. H. Electronic structure calculations of solids using the WIEN2k package for material sciences. *Computer physics communications* **2002**, *147* (1-2), 71-76.
(57) Maintz, S.; Deringer, V. L.; Tchougréeff, A. L.; Dronskowski, R. LOBSTER: A tool to extract chemical bonding from plane‐wave based DFT. **2016**.
(58) Nelson, R.; Ertural, C.; George, J.; Deringer, V. L.; Hautier, G.; Dronskowski, R. LOBSTER: Local orbital projections, atomic charges, and chemical‐bonding analysis from projector‐augmented‐wave‐based

density‐functional theory. *Journal of Computational Chemistry* **2020**, *41* (21), 1931-1940.

(59) Steinberg, S.; Dronskowski, R. The crystal orbital Hamilton population (COHP) method as a tool to visualize and analyze chemical bonding in intermetallic compounds. *Crystals* **2018**, *8* (5), 225.

(60) Müller, P. C.; Ertural, C.; Hempelmann, J.; Dronskowski, R. Crystal orbital bond index: Covalent bond orders in solids. *The Journal of Physical Chemistry C* **2021**, *125* (14), 7959-7970.

(61) Gatti, C. Chemical bonding in crystals: new directions. *Zeitschrift für Kristallographie-Crystalline Materials* **2005**, *220* (5-6), 399-457.

(62) Oyeka, E. E.; Tran, T. T. Single-Ion Behavior in New 2-D and 3-D Gadolinium $4f^7$ Materials: $CsGd(SO_4)_2$ and $Cs[Gd(H_2O)_3(SO_4)_2] \cdot H_2O$. *ACS Organic & Inorganic Au* **2022**, *2* (6), 502-510.

(63) Oyeka, E. E.; Winiarski, M. J.; Świątek, H.; Balliew, W.; McMillen, C. D.; Liang, M.; Sorolla, M.; Tran, T. T. $Ln_2(SeO_3)_2(SO_4)(H_2O)_2$ (Ln= Sm, Dy, Yb): A Mixed‐Ligand Pathway to New Lanthanide (III) Multifunctional Materials Featuring Nonlinear Optical and Magnetic Anisotropy Properties. *Angewandte Chemie International Edition* **2022**, *61* (48), e202213499.

(64) Yi, J.; Ge, X.; Liu, E.; Cai, T.; Zhao, C.; Wen, S.; Sanabria, H.; Chen, O.; Rao, A. M.; Gao, J. The correlation between phase transition and photoluminescence properties of $CsPbX_3$ (X= Cl, Br, I) perovskite nanocrystals. *Nanoscale advances* **2020**, *2* (10), 4390-4394.

(65) Perdew, J. P.; Burke, K.; Ernzerhof, M. Generalized gradient approximation made simple. *Physical review letters* **1996**, *77* (18), 3865.

(66) Giannozzi, P.; Baroni, S.; Bonini, N.; Calandra, M.; Car, R.; Cavazzoni, C.; Ceresoli, D.; Chiarotti, G. L.; Cococcioni, M.; Dabo, I. QUANTUM ESPRESSO: a modular and open-source software project for quantum simulations of materials. *Journal of physics: Condensed matter* **2009**, *21* (39), 395502.

(67) Wang, L.; Maxisch, T.; Ceder, G. Oxidation energies of transition metal oxides within the GGA+ U framework. *Physical Review B* **2006**, *73* (19), 195107.

(68) Perdew, J. P.; Ruzsinszky, A.; Csonka, G. I.; Vydrov, O. A.; Scuseria, G. E.; Constantin, L. A.; Zhou, X.; Burke, K. Restoring the density-gradient expansion for exchange in solids and surfaces. *Physical review letters* **2008**, *100* (13), 136406.

(69) Dal Corso, A. Pseudopotentials periodic table: From H to Pu. *Computational Materials Science* **2014**, *95*, 337-350.

# Supporting Information

# Inductive-Effect-Driven Tunability of Magnetism and Luminescence in Triangular Layers ANd(SO$_4$)$_2$ (A = Rb, Cs)

Xudong Huai,[1] Ebube E. Oyeka,[1] Uchenna Chinaegbomkpa,[1] Michał J. Winiarski,[2] Hugo Sanabria,[3] and Thao T. Tran*[1]

AUTHOR ADDRESS

1. Department of Chemistry, Clemson University, Clemson, South Carolina 29634, United States
2. Faculty of Applied Physics and Mathematics and Advanced Materials Center, Gdansk University of Technology, Narutowicza 11/12, 80-233 Gdansk, Poland
3. Department of Physics and Astronomy, Clemson University, Clemson, South Carolina 29634, United States

## Table of Contents

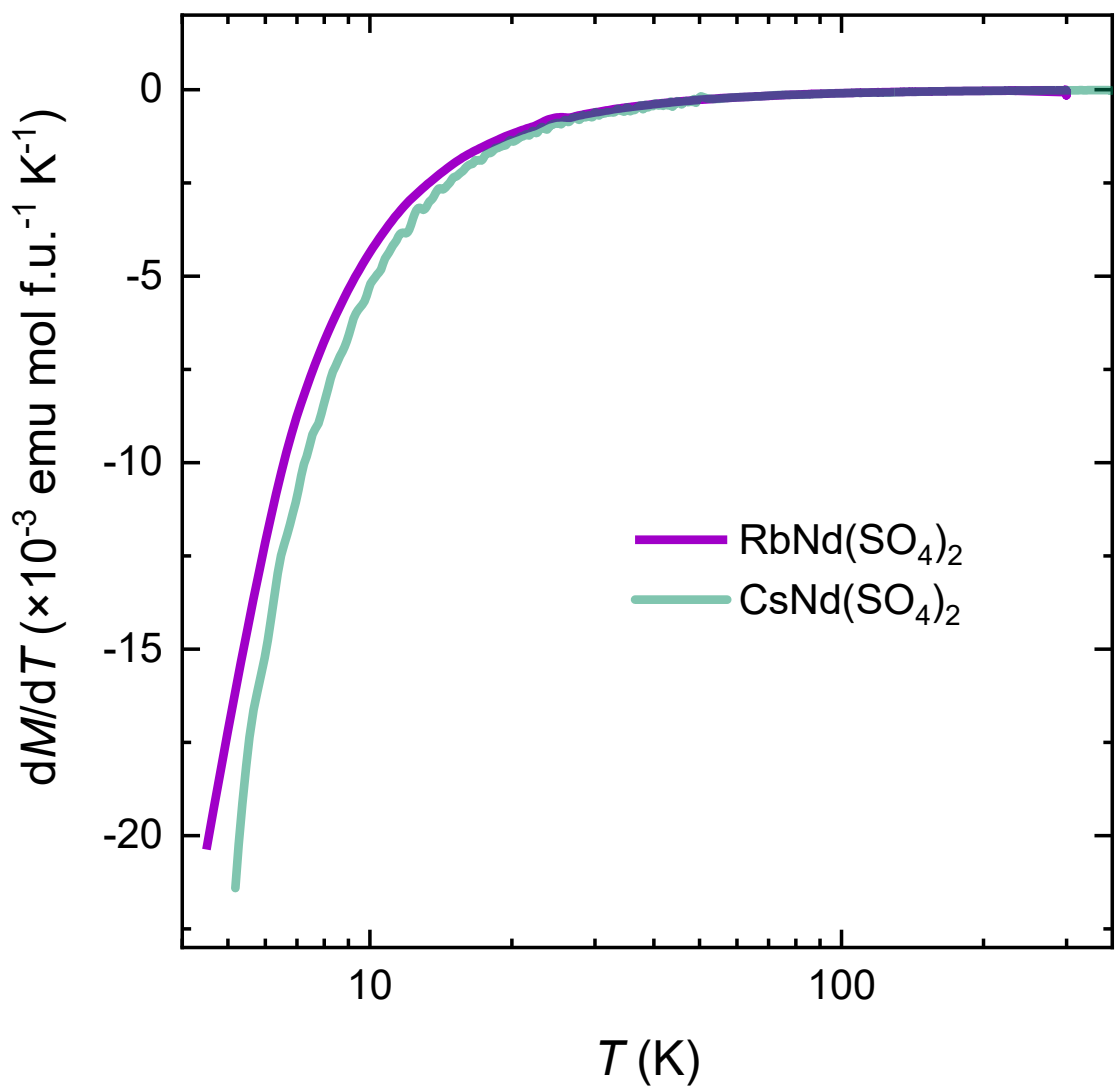

**Figure S1**: The derivative of magnetic moment as a function of ANd(SO$_4$)$_2$.

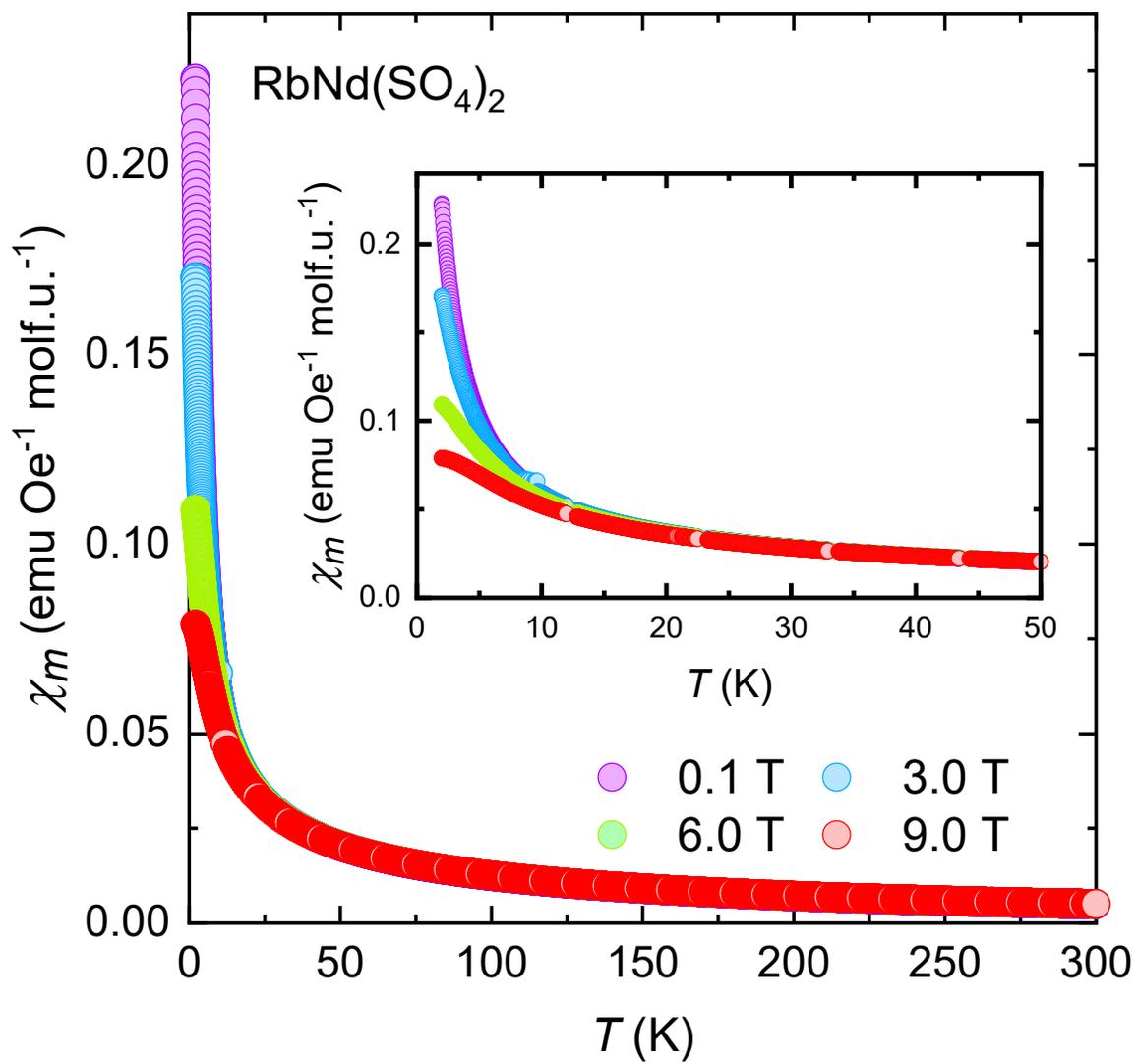

**Figure S2**: Magnetic susceptibility of RbNd(SO$_4$)$_2$. under different magnetic fields.

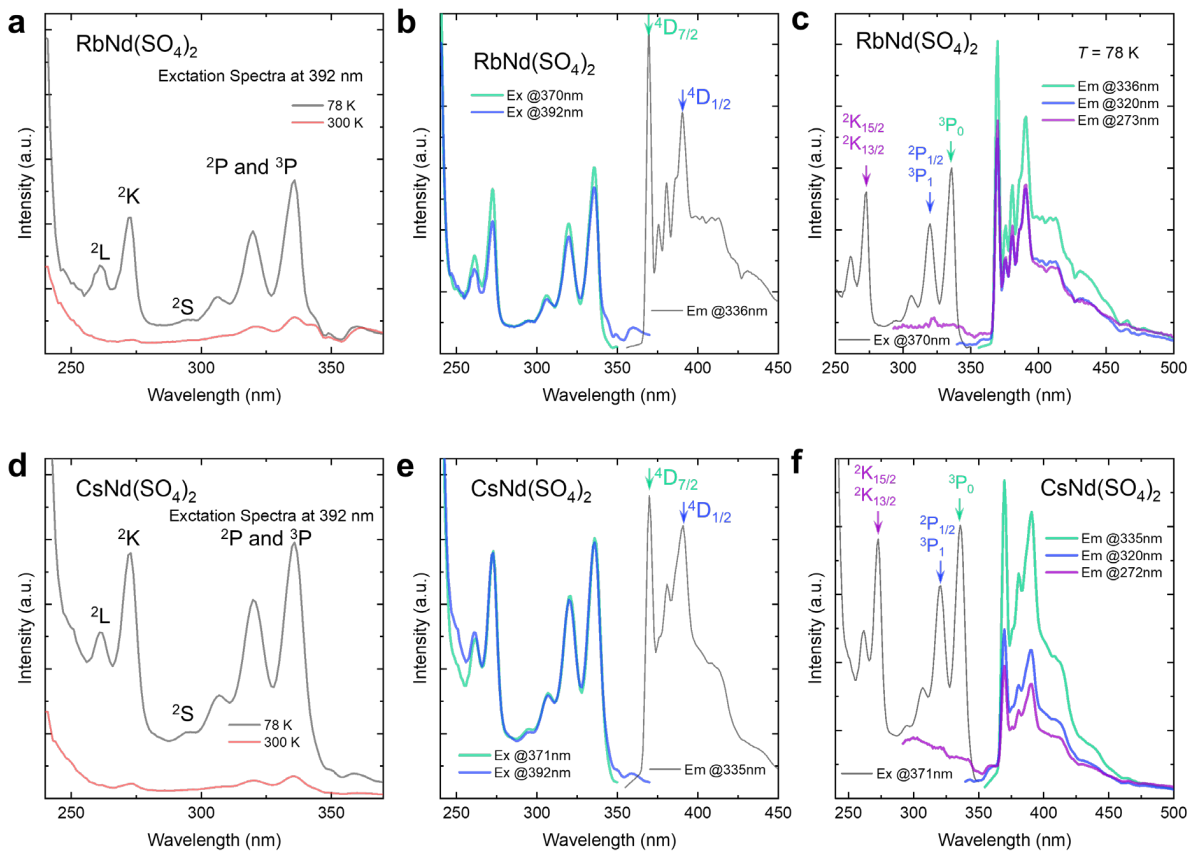

**Figure S3**: Excitation and emission spectra for ANd(SO$_4$)$_2$ (A = Rb, Cs). (a, d) The temperature dependent of the $^4I_{9/2}$ to $^3P$, $^2P$, $^2S$, $^2K$, $^2L$ excitations. (b, e) Excitation spectra of the two intense $^4D_{7/2}$, and $^4D_{1/2}$ to $^4I_{9/2}$ transitions. (c, f) Emission spectra under exaction from $^4I_{9/2}$ to $^2K$, $^2P$, and $^3P$ states.

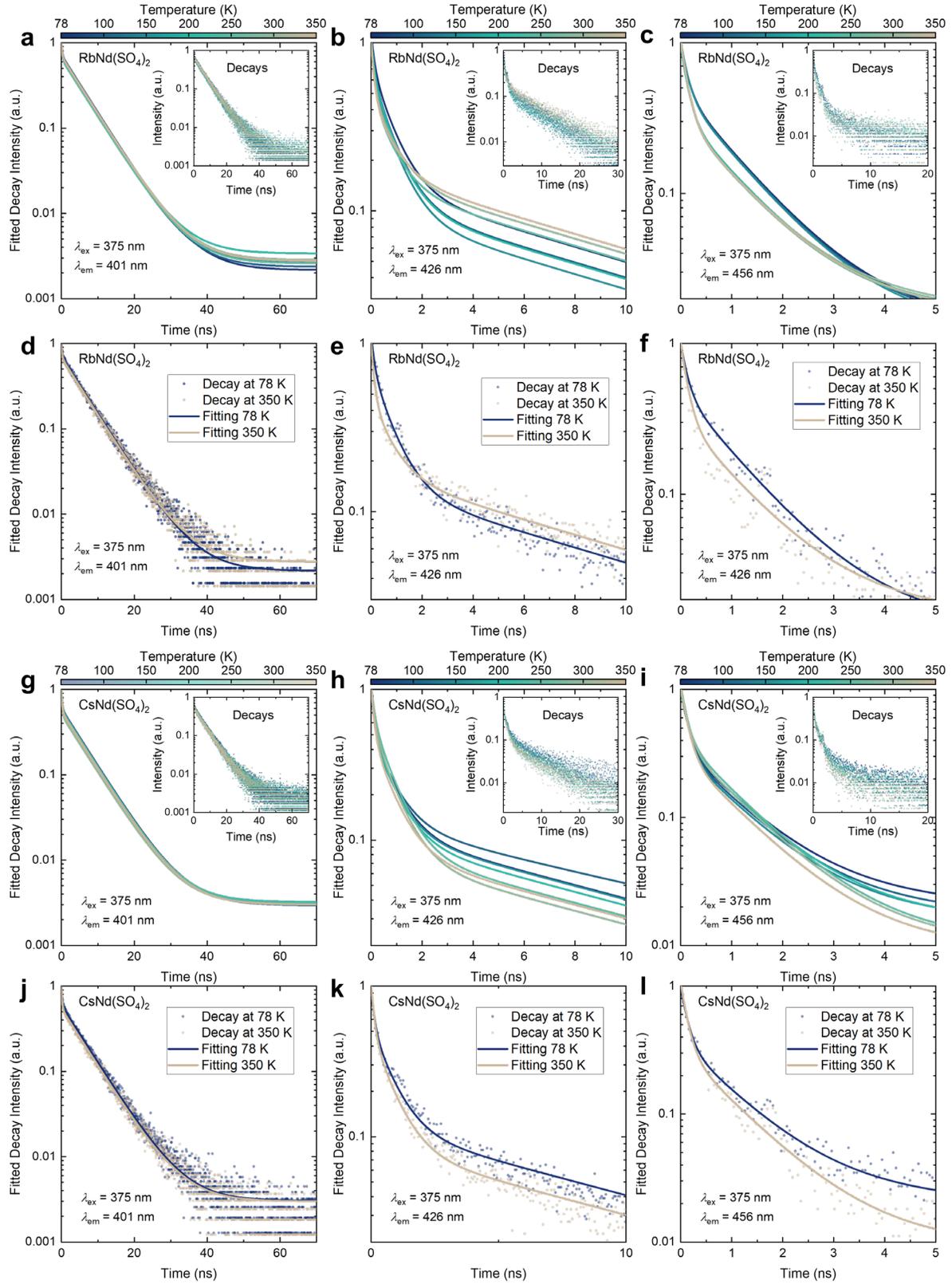

**Figure S4.** Time-resolved photoluminescence data.

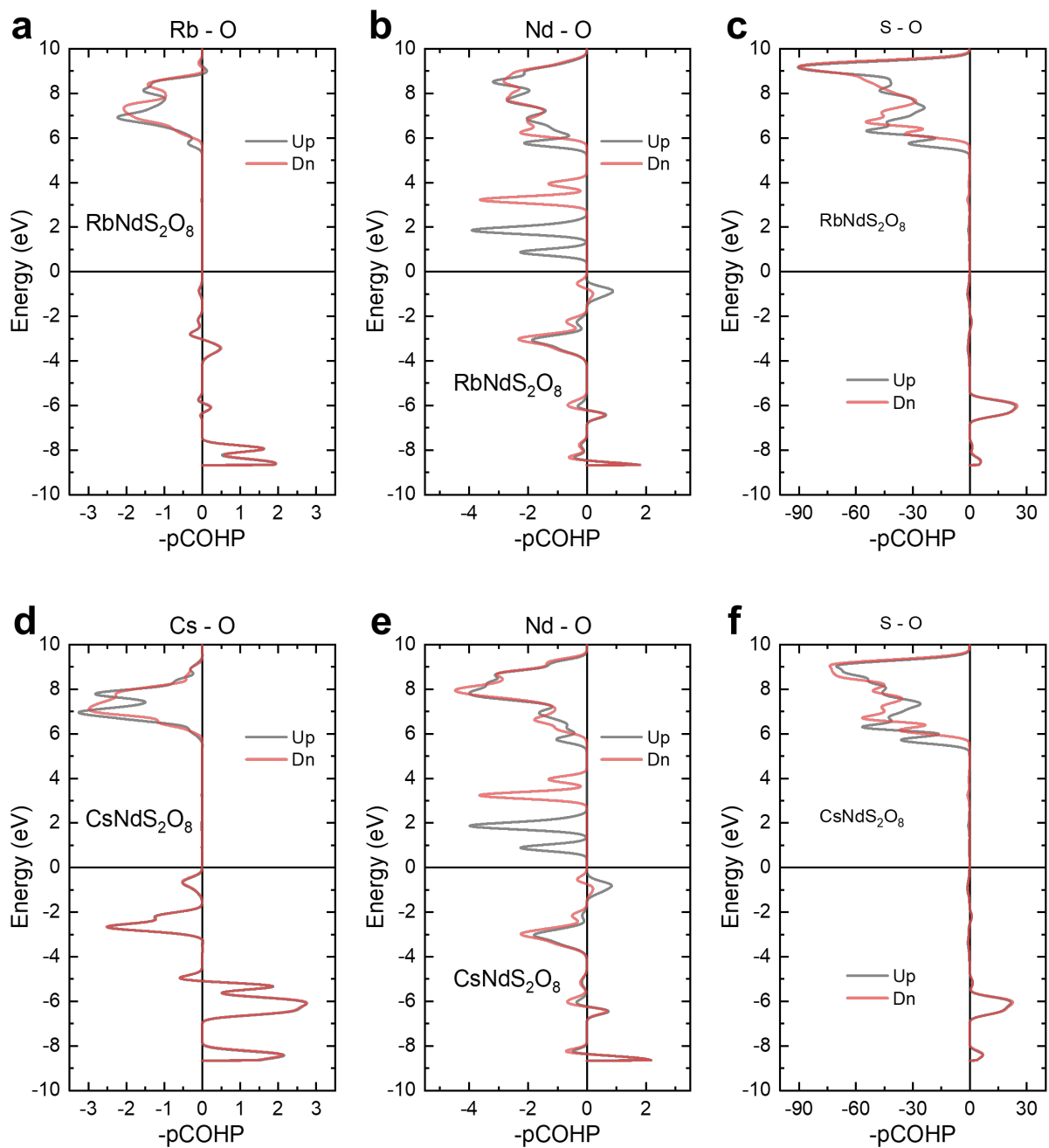

**Figure S5**: Crystal orbital Hamilton population in $ANd(SO_4)_2$ (A = Rb, Cs).

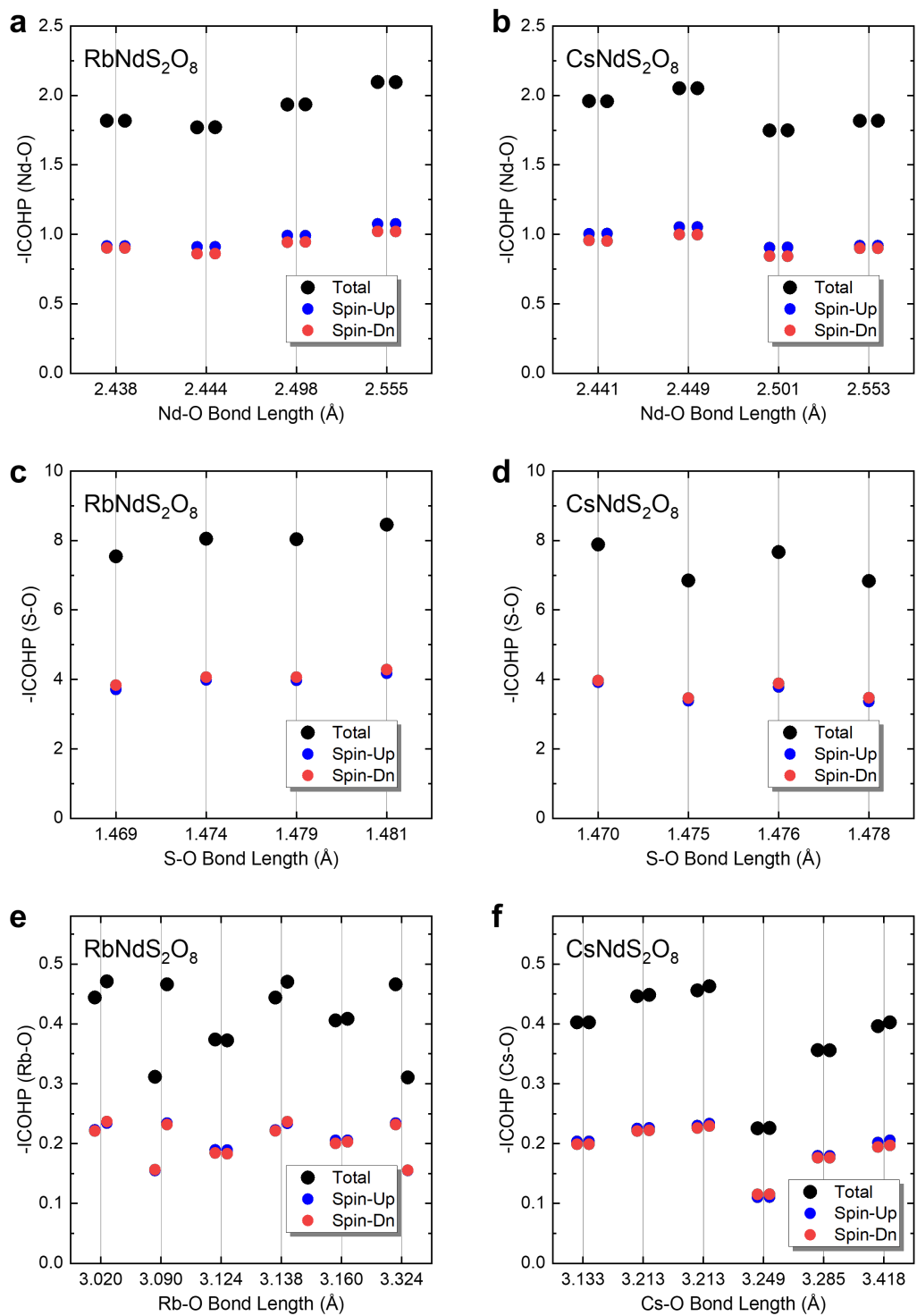

**Figure S6**: Integrated crystal orbital Hamilton population in ANd(SO$_4$)$_2$ (A = Rb, Cs).

Table S1. Single crystal X-ray diffraction data of ANd(SO$_4$)$_2$ (A = Rb, Cs).

| Empirical formula | RbNd(SO$_4$)$_2$ | CsNd(SO$_4$)$_2$ |
|---|---|---|
| Radiation Type | Mo Kα | Mo Kα |
| Formula weight | 421.83 | 469.27 |
| Crystal system | Orthorhombic | Orthorhombic |
| Space group | *Pnna* | *Pnna* |
| *a*/Å | 9.4256(11) | 9.5488(8) |
| *b*/Å | 13.3486(16) | 13.9046(11) |
| *c*/Å | 5.4229(6) | 5.4384(4) |
| Volume/Å$^3$ | 682.30(14) | 722.07(10) |
| Temperature/*K* | 298(2) | 100(2) |
| Z | 4 | 4 |
| R$_1$/% | 1.95 | 3.69 |

Table S2: Atomic positions in RbNd(SO$_4$)$_2$.

| RbNd(SO$_4$)$_2$ | x | y | z | U |
|---|---|---|---|---|
| Nd | 0.25000 | 0.00000 | 0.82051 | 0.013 |
| Rb | 0.07734 | 0.25000 | 0.25000 | 0.022 |
| S | 0.41503 | 0.09089 | 0.26393 | 0.012 |
| O 1 | 0.34050 | 0.14521 | 0.06460 | 0.016 |
| O 2 | 0.56490 | 0.12210 | 0.27540 | 0.019 |
| O 3 | 0.34750 | 0.10970 | 0.50360 | 0.018 |
| O 4 | 0.09260 | 0.01700 | 0.20180 | 0.019 |

**Table S3**: Atomic positions in CsNd(SO$_4$)$_2$.

| CsNd(SO$_4$)$_2$ | x | y | z | U |
|---|---|---|---|---|
| Nd | 0.25000 | 0.00000 | 0.82033 | 0.011 |
| Cs | 0.07811 | 0.25000 | 0.25000 | 0.011 |
| S | 0.41450 | 0.08543 | 0.26050 | 0.009 |
| O 1 | 0.56300 | 0.11400 | 0.27300 | 0.015 |
| O 2 | 0.34730 | 0.10440 | 0.50050 | 0.015 |
| O 3 | 0.09600 | 0.01740 | 0.20150 | 0.017 |
| O 4 | 0.34280 | 0.13890 | 0.06270 | 0.016 |

**Table S4**: Calculated Mulliken gross orbital population

| Atom (Orbital) | Charge in RbNd(SO$_4$)$_2$ | Charge in CsNd(SO$_4$)$_2$ |
|---|---|---|
| A (5s / 6s) | 0.14 | 0.09 |
| Nd (4f) | 3.21 | 3.21 |
| S (3p) | 2.44 | 2.44 |
| O (2p) | 5.23 | 5.28 |

**Table S5**: Magnetic properties with $Nd^{3+}$ on triangular lattice

| Compound | Nd Distance (Å) | $T_N$ (K) | $\theta_{CW}$ (K) | $\mu_{eff}$ ($\mu_B$) |
|---|---|---|---|---|
| ANd(SO$_4$)$_2$ (This work) 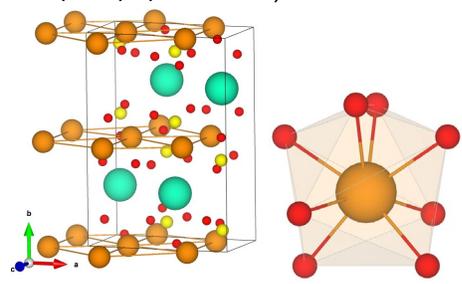 A, Nd, S, O | A = Rb: 5.099(1) 5.423(1) 5.856(1) A = Cs: 5.159(1) 5.438(1) 5.911(1) | - | A = Rb: -30.3 A = Cs: -85.6 | A = Rb: 3.6 A = Cs: 4.5 |
| NdCd$_3$P$_3$[1] 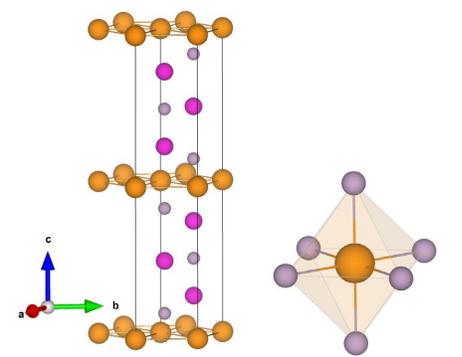 Nd, Cd, P | 4.260(1) | 0.42 | -50 | 3.62 |
| NdTa$_7$O$_{19}$[2] 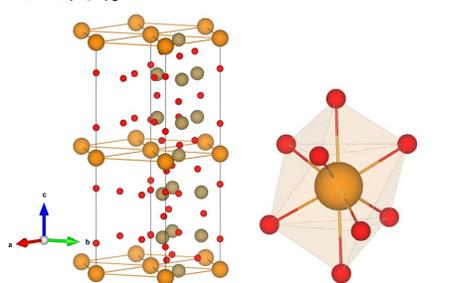 Nd, Ta, O | 6.223(1) | - | -78 | 3.8 |
| NdMgAl$_{11}$O$_{19}$[3] 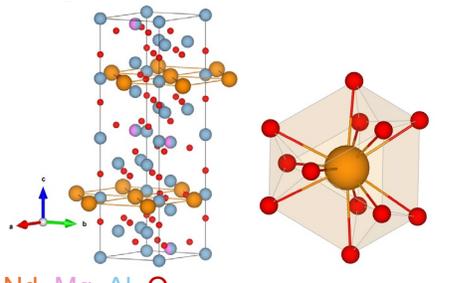 Nd, Mg, Al, O | 5.591(1) | - | -104.5 | 3.61 |
|  |  |  |  |  |
| KNdSe$_2$[4] 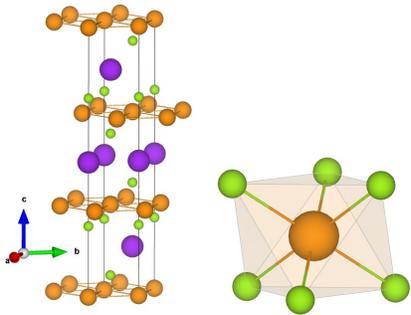 K, Nd, Se | 4.313(1) | - | H // ab: −21.35 H // c: −47.10 | H // ab: 3.24 H // c: 3.30 |
| SrNd$_2$O$_4$[5] 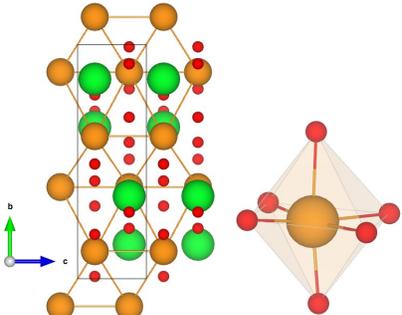 Sr, Nd, O | 3.571(1) 3.641(1) 4.127(1) | 2.28 | −39 | 3.62 |

**Table S6**: Phonon fitting results of the specific heat data.

| Variable | RbNd(SO$_4$)$_2$ | CsNd(SO$_4$)$_2$ |
|---|---|---|
| $\gamma$ | 0.02(4) | 0.01(4) |
| Number of oscillators of Einstein mode | 1.2(1) | 1.1(1) |
| Einstein temperature | 75(1) | 72(1) |
| Number of oscillators of Debye mode 1 | 4.0(2) | 3.2(3) |
| Debye temperature 1 | 280(10) | 244(14) |
| Number of oscillators of Debye mode 2 | 4.6(3) | 4.2(2) |
| Debye temperature 2 | 964(90) | 667(59) |
| $R^2$ | 0.99911 | 0.99933 |
| Total number of oscillators | 9.8(5) | 8.5(6) |

**Table S7**: Schottky effect fitting results with the specific heat data at low temperature.

| Variable | RbNd(SO$_4$)$_2$ | CsNd(SO$_4$)$_2$ |
|---|---|---|
| Schottky gap ($\Delta$)/eV at 0 T | 0(0) | 0(0) |
| States ratio ($g_0/g_1$) at 0 T | 1(0) | 1(0) |
| Schottky gap ($\Delta$)/eV at 3 T | 1.78(3) | 2.51(10) |
| States ratio ($g_0/g_1$) at 3 T | 0.27(4) | 0.97(3) |
| Schottky gap ($\Delta$)/eV at 6 T | 4.60(5) | 4.84(2) |
| States ratio ($g_0/g_1$) at 6 T | 1.61(1) | 1.52(3) |
| Schottky gap ($\Delta$)/eV at 9 T | 6.42(11) | 7.24(5) |
| States ratio ($g_0/g_1$) at 9 T | 2.22(5) | 2.93(7) |

References

1. Chamorro, J. R.; Jackson, A. R.; Watkins, A. K.; Seshadri, R.; Wilson, S. D., Magnetic order in the $S_{eff}$= 1/2 triangular-lattice compound NdCd$_3$P$_3$. *Physical Review Materials* **2023,** *7* (9), 094402.
2. Arh, T.; Sana, B.; Pregelj, M.; Khuntia, P.; Jagličić, Z.; Le, M. D.; Biswas, P. K.; Manuel, P.; Mangin-Thro, L.; Ozarowski, A., The Ising triangular-lattice antiferromagnet neodymium heptatantalate as a quantum spin liquid candidate. *Nature Materials* **2022,** *21* (4), 416-422.
3. Ashtar, M.; Gao, Y. X.; Wang, C. L.; Qiu, Y.; Tong, W.; Zou, Y. M.; Zhang, X. W.; Marwat, M. A.; Yuan, S. L.; Tian, Z. M., Synthesis, structure and magnetic properties of rare-earth REMgAl$_{11}$O$_{19}$ (RE= Pr, Nd) compounds with two-dimensional triangular lattice. *Journal of Alloys and Compounds* **2019,** *802*, 146-151.
4. Sanjeewa, L. D.; Xing, J.; Taddei, K. M.; Sefat, A. S., Synthesis, crystal structure and magnetic properties of KLnSe$_2$ (Ln= La, Ce, Pr, Nd) structures: A family of 2D triangular lattice frustrated magnets. *Journal of Solid State Chemistry* **2022,** *308*, 122917.
5. Qureshi, N.; Wildes, A. R.; Ritter, C.; Fåk, B.; Riberolles, S. X. M.; Hatnean, M. C.; Petrenko, O. A., Magnetic structure and low-temperature properties of geometrically frustrated SrNd$_2$O$_4$. *Physical Review B* **2021,** *103* (13), 134433.